\documentclass[twocolumn]{aastex631}
\usepackage{multirow}
\usepackage{makecell}
\usepackage{comment}

\begin{document}

\title{A large jet narrow-line Seyfert 1 galaxy: observations from pc to 100 kpc scales}


\correspondingauthor{Sina Chen}
\email{sina.chen@campus.technion.ac.il}

\author{Sina Chen}
\affiliation{Physics Department, Technion, Haifa 32000, Israel}

\author{Preeti Kharb}
\affiliation{National Centre for Radio Astrophysics (NCRA) - Tata Institute of Fundamental Research (TIFR), Post Bag 3, Ganeshkhind, Pune 411007, India}

\author{Silpa Sasikumar}
\affiliation{National Centre for Radio Astrophysics (NCRA) - Tata Institute of Fundamental Research (TIFR), Post Bag 3, Ganeshkhind, Pune 411007, India}
\affiliation{Departamento de Astronom\'ia, Universidad de Concepci\'on, Concepci\'on, Chile}

\author{Sumana Nandi}
\affiliation{Manipal Centre for Natural Sciences, Centre of Excellence, Manipal Academy of Higher Education, Manipal, Karnataka 576104, India}

\author{Marco Berton}
\affiliation{European Southern Observatory (ESO), Alonso de C\'ordova 3107, Casilla 19, Santiago 19001, Chile}

\author{Emilia J{\"a}rvel{\"a}}
\affiliation{Homer L.\,Dodge Department of Physics and Astronomy, The University of Oklahoma, 440 West Brooks Street, Norman, OK 73019, USA}

\author{Ari Laor}
\affiliation{Physics Department, Technion, Haifa 32000, Israel}

\author{Ehud Behar}
\affiliation{Physics Department, Technion, Haifa 32000, Israel}

\author{Luigi Foschini}
\affiliation{INAF - Osservatorio Astronomico di Brera, Via E.\,Bianchi 46, 23807, Merate (LC), Italy}

\author{Amelia Vietri}
\affiliation{Dipartimento di Fisica e Astronomia ``G.\,Galilei'', Universit{\`a} di Padova, Vicolo dell'Osservatorio 3, 35122 Padova, Italy}

\author{Minfeng Gu}
\affiliation{Key Laboratory for Research in Galaxies and Cosmology, Shanghai Astronomical Observatory, Chinese Academy of Sciences, 80 Nandan Road, Shanghai 200030, China}

\author{Giovanni La Mura}
\affiliation{INAF - Osservatorio Astronomico di Cagliari, Via della Scienza 5, 09047 Selargius (CA), Italy}
\affiliation{Laborat\'orio de Instrumenta\c{c}\~ao e F\'isica Experimental de Part\'iculas, Av.\,Prof.\,Gama Pinto 2, 1649-003 Lisboa, Portugal}

\author{Luca Crepaldi}
\affiliation{Dipartimento di Fisica e Astronomia ``G.\,Galilei'', Universit{\`a} di Padova, Vicolo dell'Osservatorio 3, 35122 Padova, Italy}
\affiliation{European Southern Observatory (ESO), Alonso de C\'ordova 3107, Casilla 19, Santiago 19001, Chile}

\author{Minhua Zhou}
\affiliation{School of Physical Science and Intelligent Education, Shangrao Normal University, 401 Zhimin Road, Shangrao 334001, China}

\begin{abstract}

We present new 1.5--8.5~GHz Very Long Baseline Array (VLBA) observations and 0.32--1.26~GHz Giant Meterwave Radio Telescope (GMRT) observations of J0354$-$1340, which is the only known `radio-quiet' (RQ) or `radio-intermediate' (RI) narrow-line Seyfert 1 galaxy with a 100-kpc two-sided radio jet.
A pc-scale one-sided jet in the southeast direction from the core emission is found in the VLBA observations, while the kpc-scale jet observed with Karl G.\,Jansky Very Large Array (VLA) and GMRT is in the south-north direction.
The core spectra on pc and kpc scales are presented in combination with the archival VLASS observations at 3.0~GHz and the VLA C configuration observations at 5.5~GHz.
The pc-scale emission dominates the kpc-scale emission above $\sim$ 5~GHz, and the spectrum is inverted due to synchrotron self-absorption.
This indicates a compact synchrotron source with a size of $\sim$ 0.04~pc, which is associated with either the jet base or the corona.
A sub-kpc scale jet, which is unresolved on scales of $\sim$ 3~arcsec, probably dominates the emission below $\sim$ 5~GHz.
Future radio observations can explore the jet structure between the pc and 100~kpc scales, the origin of their direction mismatch, and the pc-scale jet proper motion.
It remains to be explored how common such large-scale jets are in RQ or RI AGN. 

\end{abstract}


\section{Introduction}

Narrow-line Seyfert 1 galaxies (NLS1s) are a subclass of active galactic nuclei (AGN) and exhibit peculiar properties. 
Historically, NLS1s are defined by their Balmer lines from the broad-line region (BLR), in particular H$\beta$ line, having a full width at half maximum (FWHM) of $< 2000$~km\,s$^{-1}$ and a flux ratio of [O\,III]/H$\beta < 3$ \citep{Osterbrock1985,Goodrich1989}.
Compared to broad-line Seyfert 1 galaxies (BLS1s), NLS1s frequently show strong optical Fe\,II multiplets \citep{Boroson1992}, weak [O\,III] lines with blueshifted profiles \citep{Leighly2004,Boroson2005}, steep soft X-ray spectra \citep{Laor1994,Boller1996,Wang1996,Leighly1999a}, and rapid X-ray variability \citep{Pounds1995,Leighly1999b}.
However, the limit on the FWHM(H$\beta$) is somewhat arbitrary, as no clear transition is present in the distribution of properties versus FWHM(H$\beta$) at least up to 4000\,km\,s$^{-1}$ \citep[e.g.][]{Marziani2018}.

Various studies suggest that NLS1s harbor a relatively low mass central black hole (BH), typically in the range of $\log M_{\rm BH}/M_{\odot} \sim$ 6--8, compared to BLS1s with $\log M_{\rm BH}/M_{\odot} \sim$ 7--9 \citep{Peterson2011,Jarvela2015,Cracco2016,Wang2016,Chen2018}.
The narrowness of Balmer lines is the result of a low rotational velocity of the gas in the BLR around a low-mass BH.
The low BH mass generally implies a high Eddington ratio, because the luminosity of NLS1s is comparable to that of BLS1s \citep{Boroson1992,Collin2004}.
Another interpretation for the narrowness of Balmer lines is an inclination effect caused by a pole-on orientation of a disk-like BLR \citep{Decarli2008}.
However, studies of their host galaxies \citep{Boroson1992,Kollatschny2011,Jarvela2018,Olguiniglesias2020} reveal that the differences between BLS1s and NLS1s should be intrinsic rather than just an orientation effect.
Further spectropolarimetric observations can help to clarify this issue \citep[e.g.][]{Capetti2021, Sniegowska2022}.

In the radio regime, AGN can be divided into radio-loud (RL) and radio-quiet (RQ) based on the radio loudness, a ratio between the 5\,GHz radio flux density and the optical B-band flux density, $R = S_{5\rm{GHz}} / S_{4400\text{\AA}}$, larger or smaller than 10 \citep{Kellermann1989}.
The radio loudness can also be defined using the ratio of the 5\,GHz radio luminosity to the 2--10\,keV X-ray luminosity \citep{Terashima2003}, where RL objects have $\log R_{\rm X} > -3.5$ \citep{Panessa2007,Laor2008}.
The majority of NLS1s are RQ, and only a fraction ($\sim 7\%$) are RL \citep{Komossa2006}.
RL NLS1s are likely to harbor relativistic jets, which are the predominant origin of the radio emission, with a few exception \citep[e.g.,][]{Caccianiga2015}.
Radio observations of RL NLS1s with the Very Long Baseline Array (VLBA) show that they exhibit pc-scale jets, compact morphologies, flat or inverted spectra, high brightness temperatures, and significant polarization \citep{Abdo2009a,Abdo2009b,Gu2015,Lister2018}.
In some cases, apparent superluminal motion is observed as well \citep{Lister2013,Lister2016,Lister2019}.
However, on kpc scales, NLS1s rarely exhibit extended radio structures and generally show an unresolved component \citep{Berton2018,Chen2020,Jarvela2022}.
In RQ NLS1s, the origin of the radio emission is still an open question.
Various radio emission mechanisms may be involved, including star formation (SF), AGN-driven wide-angle winds interacting with ambient medium, collimated low-power jets, accretion disk coronal emission, and free-free emission from AGN photoionized gas \citep[see][for a review]{Panessa2019}.
Currently, it is often hard to distinguish the radio emission from these different emission processes, in particular when the radio features are unresolved or marginally resolved.

In a previous radio survey with the Karl G.\,Jansky Very Large Array (VLA) C configuration at 5.5\,GHz, we found a particularly interesting RQ NLS1 6dF~J0354328$-$134008 (hereafter J0354$-$1340), which shows a $\sim$ 200\,kpc core-jet/lobe structure \citep{Chen2020} resembling a Fanaroff-Riley type II (FR\,II) radio galaxy.
At present, kpc-scale to Mpc-scale jets are common to see in RL AGN.
When $R \sim$ a few tens, which is close to the boundary between RL and RQ, radio structures with sizes of typically a few tens kpc are also usually seen \citep[e.g.\,][]{Jarvis2021}.
However, such a large jet ($\sim$ 100\,kpc) in a RQ AGN, which is generally not expected to harbor a relativistic jet, is unprecedented.
This raises questions why the radio luminosity of this RQ NLS1 is low despite the presence of a radio jet extending to $\sim$ 100\,kpc, and whether there are differences in the radio jets between RL and RQ objects.
J0354$-$1340 is thus an ideal laboratory to study the difference in the radio jets between RL and RQ AGN.

In this work, we report our new observations with the VLBA in L (18\,cm), C (6\,cm), and X (3.6\,cm) bands, and the upgraded Giant Meterwave Radio Telescope (GMRT) observations in Band 3 (320\,MHz), Band 4 (650\,MHz), and Band 5 (1.26\,GHz) of this object.
We aim to characterize the core region which is unresolved in our previous VLA observations, and obtain the radio spectra, which carry diagnostic information on the emission mechanism, on both pc and kpc scales.

This paper is organized as follows. In Section~\ref{target_section} we describe the previous studies on this target, and in Section~\ref{data_section} the data reduction of our new VLBA and uGMRT observations is presented. The results are presented in Section~\ref{result_section} and discussed in Section~\ref{discuss_section}. The summary is given in Section~\ref{summary_section}.
Throughout this work, we use the flux density and spectral index convention of $S_{\nu} \propto \nu^{\alpha}$.
We adopt a standard $\Lambda$CDM cosmology with a Hubble constant $H_0$ = 70~km\,s$^{-1}$\,Mpc$^{-1}$, $\Omega_{\Lambda}$ = 0.73 and $\Omega_{\rm M}$ = 0.27 \citep{Komatsu2011}.

\section{Target} \label{target_section}

J0354$-$1340 ($z$ = 0.076, corresponding to 1.488 pc\,mas$^{-1}$ or kpc\,arcsec$^{-1}$) was classified as a NLS1 based on the optical spectrum from the Six-degree Field Galaxy Survey (6dFGS) \citep{Jones2009} with a FWHM of the broad H$\beta$ line of $\sim$ 2100\,km\,s$^{-1}$ \citep{Chen2018}.
We estimated the BH mass based on the FWHM and the luminosity of H$\beta$ line \citep{Greene2005b}, the Eddington ratio, and the flux of Fe\,II ($\lambda$4570) following the Section 3 in \citet{Chen2022a}.
The object has an [O\,III]/H$\beta$ flux ratio of 0.2, an Fe\,II/H$\beta$ flux ratio of 1.0, a BH mass of $\log M_{\rm BH}/M_{\odot}$ = 7.1 with a systematic uncertainty of 0.5~dex, and an Eddington ratio of $L/L_{\rm Edd}$ = 0.6, all of which are typical for NLS1s \citep{Boroson1992}.

The object falls in the RQ \citep{Kellermann1989} or radio-intermediate (RI) regime \citep[$3<R<200$;][]{Falcke1996}.
The optical radio loudness is $R \sim 7$, where the radio flux at 5~GHz includes both the core and the lobes in the VLA C-configuration observations (see Table~\ref{flux_gmrt}), and the optical flux at B-band 4400{\AA} is measured from the 6dFGS spectrum with the subtraction of the host galaxy \citep{Chen2018}.
The optical radio loudness of this object is close to the somewhat arbitrary boundary of $R=10$ between RL and RQ sources \citep{Kellermann1989}.
The object was also detected in X-ray, with a flux of $2.6 \times 10^{-12}$~erg\,s$^{-1}$\,cm$^{-2}$ at 0.2--12~keV from the Second {\it XMM-Newton} Slew Survey Catalogue \citep{Saxton2008}, and a flux of $1.9 \times 10^{-12}$~erg\,s$^{-1}$\,cm$^{-2}$ at 0.1--2.4~keV from the second ROSAT allsky survey source catalog \citep{Boller2016}.
The X-ray radio loudness is $\log R_{\rm X} = -4.0$, which is computed using the same radio flux in the calculation of $R$, and the X-ray flux at 0.2--12~keV.
The $R_{\rm X}$ is on the high end of RQ AGN that cluster around $\log R_{\rm X} \simeq -5.0$ \citep{Laor2008}.
We note that there are possible uncertainties, due to variability in non-simultaneous observations, and aperture effects in the radio, optical, and X-ray bands.

The object exhibits an extended radio morphology with a projected size of $\sim$ 180\,kpc in the VLA C configuration observations at 5.5\,GHz \citep[see Figure A1 in][]{Chen2020}.
Based on the jet and counter-jet flux ratio, an assumption of a jet speed of $0.5c$ gives a viewing angle of about 47$^{\circ}$ and a deprojected size of about 240\,kpc \citep{Vietri2022}, which suggests that the object is not viewed pole-on, and may exclude the possibility of an underestimated BH mass due to a face-on flat BLR.
The central core is unresolved on a resolution of 3.5\,arcsec ($\sim 5$\,kpc) and has a flux density of $S_{5.5}$ = 5.1~mJy. 
The southern and northern lobes have projected lengths of 93.5\,kpc (62.2\,arcsec) and 83.9\,kpc (55.7\,arcsec) from the central core respectively, which are similar to the jet sizes in FR\,II radio galaxies \citep{Hardcastle1998,Kharb2008}.

The object was also detected with the NRAO VLA Sky Survey (NVSS) at 1.4\,GHz (VLA D configuration) with a flux density of $S_{1.4}$ = 14.9~mJy on a resolution of 45\,arcsec ($\sim$ 65\,kpc).
The NVSS map displays an elongated structure while the two-sided lobes are marginally resolved \citep[see Figure A2 in][]{Chen2020}.
This morphology is consistent with what is observed with the VLA C configuration. In the 5.5\,GHz image, the radio lobes are resolved, and the central core is likely dominated by the compact jet base emission with a flat in-band spectral slope of $\alpha_{5-6} = -0.12$. In the 1.4\,GHz image, the radio lobes are unresolved, thus the radio emission from both the core and the lobes is included.
The jet radiative and kinetic power was estimated based on the 1.4\,GHz flux and the 5.5\,GHz core flux following two methods in \citet{Cavagnolo2010} and \citet{Foschini2014}, respectively. We obtained $P_{\rm jet} \sim 10^{43}$~erg\,s$^{-1}$, which is in the low-power tail of the distribution of RL NLS1s ($\sim 10^{42.5}-10^{45.6}$~erg\,s$^{-1}$) \citep{Foschini2015}.

J0354$-$1340 is hosted in a barred spiral galaxy as found in a photometric decomposition study of its host galaxy \citep{Vietri2022}, which is consistent with NLS1s preferably hosted in disk-like galaxies, often with pseudo-bulges and bars \citep{Crenshaw2003b,Deo2006,Orbandexivry2011,Mathur2012,Jarvela2018,Olguiniglesias2020}.
Traditionally, powerful relativistic jets are harbored in early-type elliptical galaxies with massive BHs \citep[e.g.,][]{Urry1995,Laor2000,Best2005}.
However, further studies have found that large radio jets can also be launched in late-type spiral galaxies with low-mass BHs \citep{Hota2011,Bagchi2014,Webster2021a,Webster2021b}, which changes our understanding.

\section{Observations and data reduction} \label{data_section}

\subsection{VLBA}

The VLBA observations were carried out in L (18\,cm), C (6\,cm), and X (3.6\,cm) bands using 10 main VLBA stations in November 2021 (Program ID: BC271).
We used the digital down conversion observing system with a 2-bit sampling at a data rate of 2\,Gbps and four intermediate frequency (IF) bands with dual polarization and 128 MHz bandwidth.
The central frequencies are 1.5\,GHz in L band, 5.5\,GHz in C band, and 8.5\,GHz in X band.
Phase-referencing continuum observations were performed.
The observation was five hours long with scans at different frequencies interleaved, which can yield a better UV coverage.
A five-minute nodding cycle was used with three minutes on a target and one minute on a phase calibrator (J0351$-$1153) before and after the target.
A ``Fringe Finder'' (J0530+1331) was observed three times when all the antennas were up.
This strategy yields an integration time of about 50 minutes on the target at each frequency.

The data was calibrated using the VLBA data calibration pipeline procedure VLBARUN \footnote{\url{http://www.aips.nrao.edu/vlbarun.shtml}} in the Astronomical Image Processing System \citep[AIPS;][]{Greisen2003}.
A step-by-step usage of VLBARUN is described in Appendix C of the AIPS Cookbook \footnote{\url{http://www.aips.nrao.edu/cook.html}}.
The standard steps include first correcting the visibility for ionospheric delays and Earth Orientation Parameters, then applying amplitude corrections from digital sampling, finding the phase delays using the fringe finder and applying phase solutions.
An inspection of all baselines (pairs of antennas) for radio frequency interference (RFI) was performed. Data suffering from RFI was removed using the AIPS task UVFLG.
A self-calibration was applied to the calibrators.
Once the calibration procedure was completed, we applied the final calibration table on the target and split the data of the target using the AIPS task SPLIT.

\begin{figure*}[ht!]
\includegraphics[width=2\columnwidth, trim={0cm, 3.8cm, 0cm, 2.3cm}, clip]{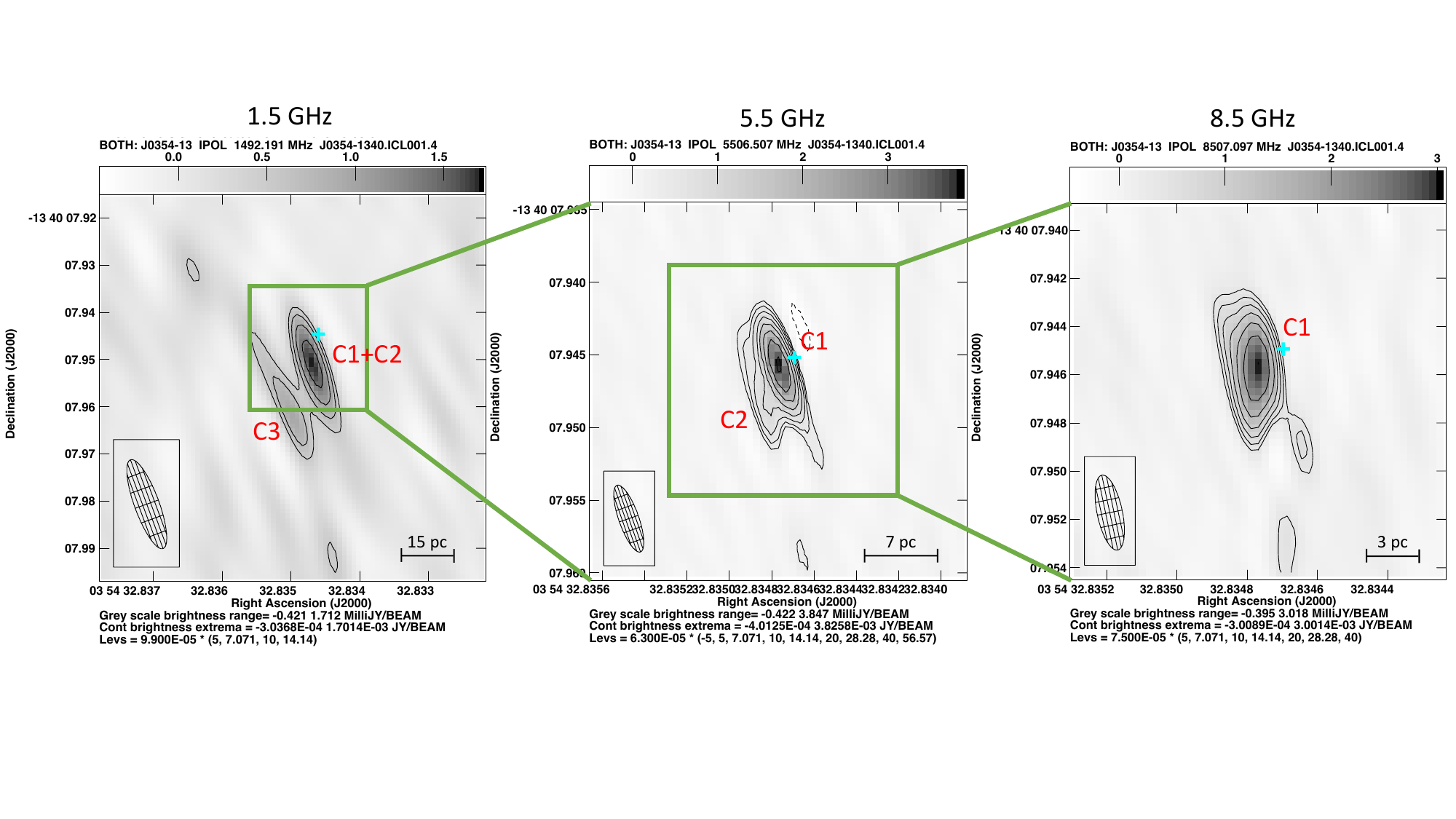}
\caption{VLBA maps at 1.5 (left), 5.5 (middle), and 8.5 (right) GHz.
Background noises $\sigma$ are reported in Table~\ref{flux_vlba}. The contour levels are $\sigma \times (-5, 5 \times \sqrt{2}^n)$ where $n \in [0,10]$. The size and orientation of the synthesized beam are shown in the lower-left corner. The grey scale indicates the intensity in a unit of Jy\,beam$^{-1}$ in a linear scale. The cyan cross marks the {\it Gaia} position. The object has three components, which are labeled as C1, C2, and C3.}
\label{map_vlba}
\end{figure*}

\begin{figure}[ht!]
\includegraphics[width=\columnwidth, trim={8cm, 0cm, 8cm, 0cm}, clip]{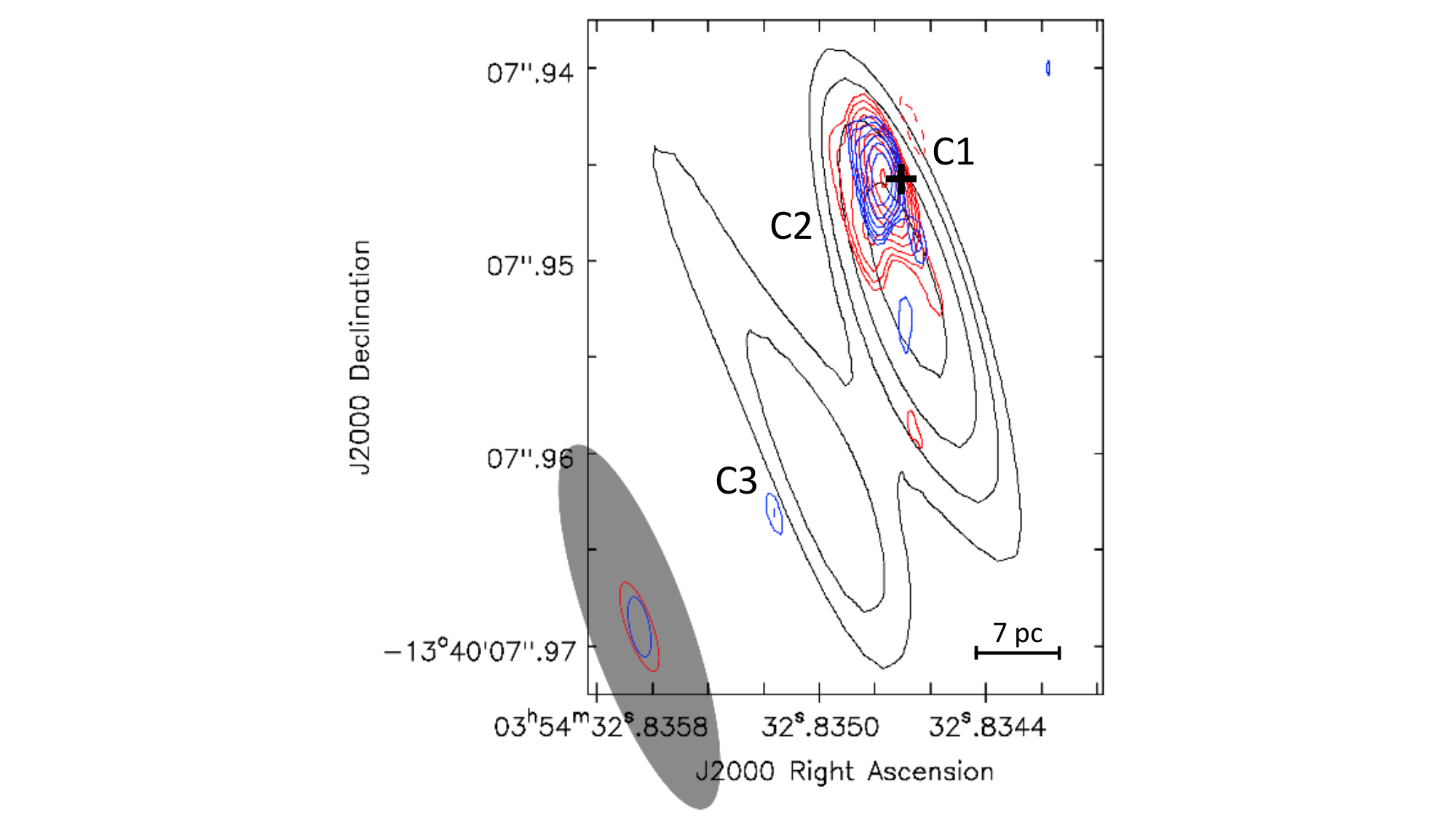}
\caption{An overlap of VLBA maps at 1.5\,GHz (black), 5.5\,GHz (red), and 8.5\,GHz (blue). The contour levels are the same as in Figure~\ref{map_vlba}. The size and orientation of the synthesized beams are shown in the lower-left corner. The black cross represents the {\it Gaia} position. The extension of C2 and C3 to the southeast of the core C1 represents the pc-scale jet direction.}
\label{overlap}
\end{figure}

We imaged the visibility data of the target using the AIPS task IMAGR.
This task uses the CLEAN algorithm to deconvolve the ``dirty image'' with the point spread function (PSF), to obtain the residual and the ``clean image''.
We chose `briggs' weighting (i.e.\ robust=0) which balances the sensitivity and the angular resolution.
The source images were not self-calibrated.
In order to obtain comparable resolutions in L, C, and X bands, we created tapered images via setting a UV range of 6000--50000\,k$\lambda$ in three bands.
The lower and upper limits are equivalent to the minimum UV range in X band and the maximum UV range in L band.
In the tapering, we used `natural' weighting (i.e.\ robust=5), which maximizes the sensitivity at the expense of the angular resolution.
The final images, as shown in Figure~\ref{map_vlba}, were inspected using the AIPS task IMEAN and IMFIT, to obtain the peak and total flux densities, the background noise, and the source size, which are reported in Tables \ref{size_vlba} and \ref{flux_vlba}.
We use a 5$\sigma$ level as the detection criterion.
We further overlap the contour maps in all bands, as shown in Figure~\ref{overlap}, to see how the emission aligns at different frequencies and on different scales.

\begin{table*}[ht!]
\caption{The beam and source sizes in the VLBA observations. Columns:
(1) frequency,
(2) right ascension,
(3) declination,
(4) distance between the VLBA and Gaia positions,
(5) major axis of the beam, 
(6) minor axis of the beam, 
(7) position angle of the beam, 
(8) major axis of the source, 
(9) minor axis of the source, 
(10) position angle of the source,
(11) deconvolved major axis of the source, 
(12) deconvolved minor axis of the source, 
(13) deconvolved position angle of the source.}
\label{size_vlba}
\centering
\scriptsize
\begin{tabular}{ccccccccccccc}
\hline
\hline
Frequency & \multicolumn{2}{c}{Coordinates} & Offset & \multicolumn{3}{c}{Beam size} & \multicolumn{3}{c}{Source size} & \multicolumn{3}{c}{Deconvolved source size} \\
$\nu$ & RA & Dec & $d_{\rm Gaia}$ & $\theta_{\rm maj}$ & $\theta_{\rm min}$ & PA & $\theta_{\rm maj}$ & $\theta_{\rm min}$ & PA & $\theta_{\rm maj}$ & $\theta_{\rm min}$ & PA \\
(GHz) & (hh:mm:ss) & (dd:mm:ss) & (mas) & (mas) & (mas) & (degree) & (mas) & (mas) & (degree) & (mas) & (mas) & (degree) \\
(1) & (2) & (3) & (4) & (5) & (6) & (7) & (8) & (9) & (10) & (11) & (12) & (13) \\
\hline
1.5 C1+C2 & 03:54:32.8347 & $-$13:40:07.9517 & 6.5 & \multirow{2}{*}{20.03} & \multirow{2}{*}{5.20} & \multirow{2}{*}{20.2} & 20.95 & 4.88 & 18.7 & $<$10.02 & $<$2.60 & - \\
1.5 C3 & 03:54:32.8351 & $-$13:40:07.9593 & 14.9 & & & & 31.96 & 4.82 & 25.8 & 25.02 & $<$2.60 & - \\
\hline
5.5 C1 & 03:54:32.8348 & $-$13:40:07.9458 & 0.8 & \multirow{2}{*}{4.86} & \multirow{2}{*}{1.37} & \multirow{2}{*}{19.4} & 5.03 & 1.39 & 18.7 & $<$2.43 & $<$0.69 & - \\
5.5 C2 & 03:54:32.8348 & $-$13:40:07.9479 & 3.2 & & & & 6.66 & 1.40 & 17.0 & 4.57 & $<$0.69 & - \\
\hline
8.5 C1 & 03:54:32.8348 & $-$13:40:07.9457 & 0.9 & 3.17 & 1.05 & 12.0 & 3.76 & 1.52 & 5.5 & 2.11 & 0.89 & 168.9 \\
\hline
\end{tabular}
\end{table*}

\begin{table*}[ht!]
\caption{The flux densities at 1.5, 5.5, and 8.5 GHz in the VLBA observations. Columns:
(1) frequency, 
(2) peak flux density of the full resolution map, 
(3) total flux density of the full resolution map, 
(4) background noise of the full resolution map, 
(5) logarithm of surface brightness temperature in unit of K,
(6) peak flux density of the tapered map, 
(7) total flux density of the tapered map, 
(8) background noise of the tapered map,
(9) spectral slope.
The tapered maps have a UV range of 6000--50000\,$k\lambda$ in the three bands.}
\label{flux_vlba}
\centering
\scriptsize
\begin{tabular}{ccccccccc}
\hline\hline
Frequency & \multicolumn{3}{c}{Full resolution maps} & Brightness & \multicolumn{3}{c}{Tapered (6000--50000 $k\lambda$) maps} & Spectral slope \\
$\nu$ & $S_{\rm peak}$ & $S_{\rm total}$ & RMS & $\log T_{\rm B}$ & $S_{\rm peak}$ & $S_{\rm total}$ & RMS & $\alpha$ \\
(GHz) & (mJy\,beam$^{-1}$) & (mJy) & (mJy\,beam$^{-1}$) & - & (mJy\,beam$^{-1}$) & (mJy) & (mJy\,beam$^{-1}$) & - \\
(1) & (2) & (3) & (4) & (5) & (6) & (7) & (8) & (9) \\
\hline
1.5 C1+C2 & 1.73 $\pm$ 0.07 & 1.73 $\pm$ 0.07 & \multirow{2}{*}{0.099} & 7.8 & \multirow{2}{*}{1.47 $\pm$ 0.07} & \multirow{2}{*}{1.47 $\pm$ 0.07} & \multirow{2}{*}{0.096} & \multirow{5}{*}{\makecell{$\alpha_{1-5} = 0.93 \pm 0.04$ \\ $\alpha_{5-8} = 0.17 \pm 0.05$}} \\
1.5 C3 & 0.92 $\pm$ 0.07 & 1.36 $\pm$ 0.16 & & 7.3 & & & & \\
\cline{1-8}
5.5 C1 & 3.85 $\pm$ 0.04 & 4.07 $\pm$ 0.07 & \multirow{2}{*}{0.063} & 8.2 & \multirow{2}{*}{4.72 $\pm$ 0.03} & \multirow{2}{*}{4.93 $\pm$ 0.06} & \multirow{2}{*}{0.074} & \\
5.5 C2 & 0.90 $\pm$ 0.04 & 1.26 $\pm$ 0.09 & & 7.4 & & & & \\
\cline{1-8}
8.5 C1 & 3.01 $\pm$ 0.05 & 5.17 $\pm$ 0.13 & 0.075 & 7.9 & 4.70 $\pm$ 0.05 & 5.30 $\pm$ 0.10 & 0.094 & \\
\hline
\end{tabular}
\end{table*}

\subsection{GMRT}

The GMRT observations at 320~MHz (Band 3), 650~MHz (Band 4), and 1.26~GHz (Band 5) were carried out in January 2021 (Program ID: 39\_006).
We used the full array of 30 antennas.
The total observing time is 13 hours with 6 hours in Band 3 and 3.5 hours in both Bands 4 and 5.
The amplitude calibrator (3C147) was observed for 8 minutes at the start and end of each run, and the phase calibrator (0405$-$131) was observed for 5 minutes every 30 minutes on the target.
This yields an on-source time of about 4 hours in Band 3 and 2 hours in both Bands 4 and 5.

The Band-4 and Band-5 broad-band data was calibrated using a Common Astronomy Software Applications (CASA) based pipeline \citep{Ishwara2020} \footnote{\url{http://www.ncra.tifr.res.in/~ishwar/pipeline.html}}.
The pipeline performs two cycles of flagging and calibration on the data.
The central 80-90\% channels were used to keep the bandwidth smearing negligible.
The imaging of the science target was done using the CASA task TCLEAN.
Four rounds of phase-only self-calibration and four rounds of amplitude and phase self-calibration with solution intervals of 8, 4, 2, and 1 minutes were carried out.
Flagging was also performed before each self-calibration round.
The pipeline produces eight images, the image quality improves in the third and fourth cycles of self-calibrations.
Details of the pipeline procedure can be seen in \citet{Ishwara2020}.

The angular resolutions are about 4~arcsec at 650~MHz and 2~arcsec at 1.26~GHz.
The object also has archival observations with the VLA Sky Survey (VLASS) at 3.0~GHz and the VLA C configuration at 5.5~GHz, which have resolutions of 2.5~arcsec and 3.5~arcsec respectively.
In order to build a kpc-scale spectrum less biased to the resolution, we further created a tapered map in Band 5 via setting a UV range of 40k$\lambda$, which corresponds to an angular resolution of 3~arcsec.
We use the tapered map, instead of the full resolution map, at 1.26~GHz in this study hereafter.

The Band-3 narrow-band data was reduced using the Source Peeling and Atmospheric Modeling \citep[SPAM;][]{Intema2009,Intema2014a,Intema2014b}, which is an AIPS based Python package that provides semi-automated data reduction scripts for all sub-GHz frequencies at GMRT \footnote{\url{https://www.intema.nl/doku.php?id=huibintemaspampipeline}}. The flux density and bandpass calibration was performed using the calibrator 3C147. In the main SPAM pipeline, the pre-calibrated target visibility data was processed and the final image with primary beam correction is produced. Since the Band-3 broad-band data probably was strongly affected by RFI and/or ionospheric phase errors, the lack of a direction-dependent ionospheric phase calibration, which is currently not available in any other software for the GMRT broad-band data, produces noisy images that are not usable for the analysis.

\begin{figure*}[ht!]
\centering
\includegraphics[width=18cm, trim={0cm, 2cm, 0cm, 1.7cm}, clip]{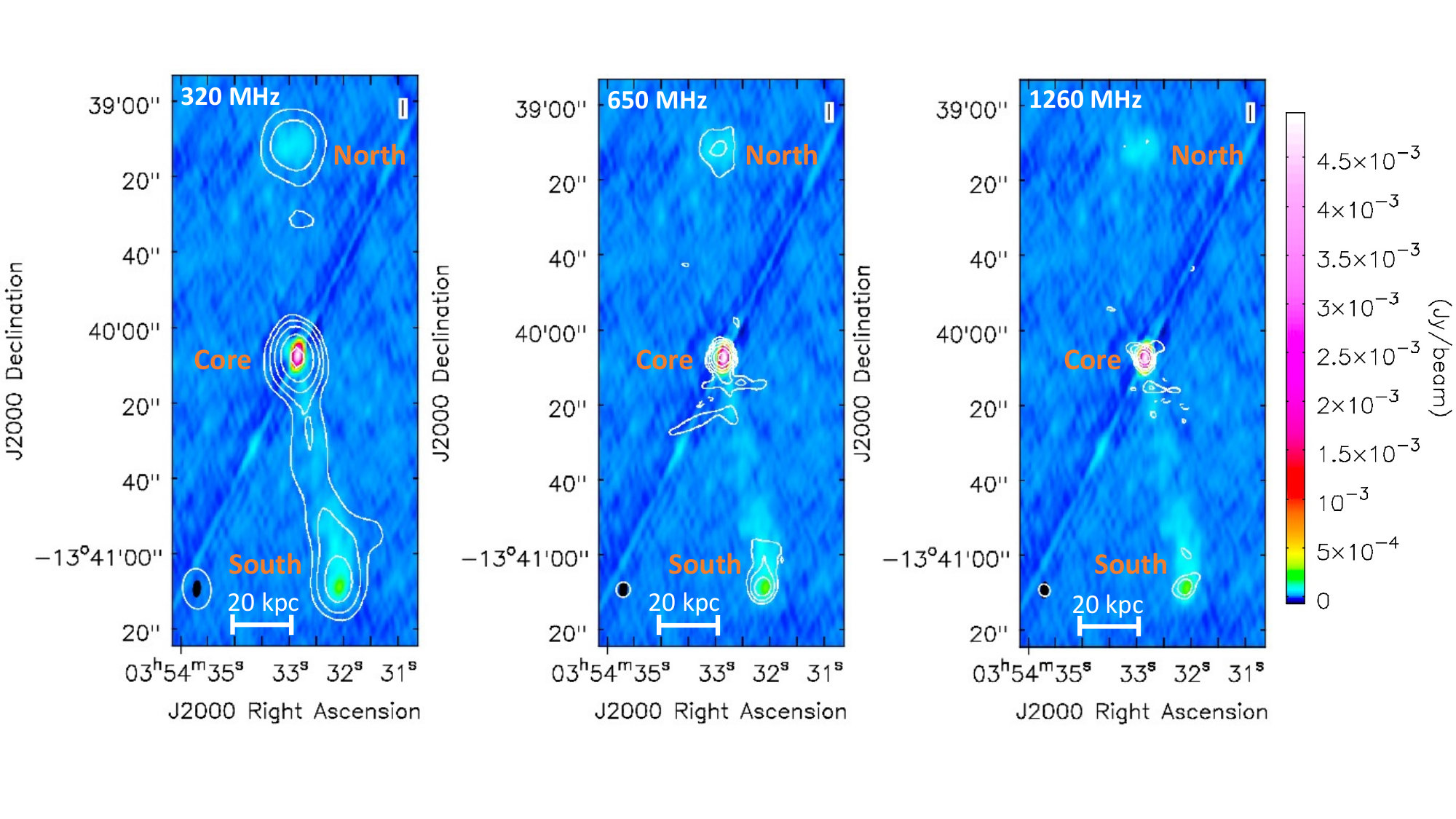}
\caption{The GMRT maps at 320~MHz (left), 650~MHz (middle), and 1.26~GHz (right) in white contours, overlaid on the color VLA (C configuration) map at 5.5~GHz.
Background noises $\sigma$ are reported in Table~\ref{flux_gmrt}.
The contour levels are $\sigma \times (-5, 5 \times 2^n)$ where $n \in [0,5]$. The size and orientation of the synthesized beams are shown in the lower-left corner of each panel (black filled ellipse for VLA and white solid line for GMRT).
The color bar on the right indicates the intensity in units of Jy\,beam$^{-1}$ in a linear scale. The object exhibits three components in both radio arrays, a core and southern and northern lobes.}
\label{map_gmrt}
\end{figure*}

We choose the image with the highest signal-to-noise (S/N) ratio in this study hereafter.
The images were inspected using the CASA task IMVIEW.
The background noise is measured in a source-free region nearby the target.
We modeled the source with a Gaussian fit on the image plane and deconvolved it from the beam, to measure the position, size, position angle, and the integrated and peak flux densities.
Figure~\ref{map_gmrt} shows the GMRT maps at 650~MHz and 1.26~GHz, which are presented in white contours, overlaid on the VLA (C configuration) map at 5.5~GHz from \citet{Chen2020}, which is presented in color.
Tables \ref{size_gmrt} and \ref{flux_gmrt} report the source size, the total and peak flux densities at 0.65 and 1.26 GHz from GMRT, at 3.0~GHz from VLASS, and at 5.5~GHz from VLA C configuration.
We use a 5$\sigma$ level as the detection criterion.

\begin{table*}[ht!]
\caption{The beam and source sizes in the GMRT, VLASS, and VLA observations. Columns:
(1) telescope,
(2) frequency,
(3) component,
(4) major axis of the beam,
(5) minor axis of the beam,
(6) position angle of the beam,
(7) major axis of the source,
(8) minor axis of the source,
(9) position angle of the source,
(10) deconvolved major axis of the source,
(11) deconvolved minor axis of the source,
(12) deconvolved position angle of the source.}
\label{size_gmrt}
\centering
\scriptsize
\begin{tabular}{cccccccccccc}
\hline
\hline
Telescope & Frequency & Component & \multicolumn{3}{c}{Beam size} & \multicolumn{3}{c}{Source size} & \multicolumn{3}{c}{Deconvolved source size} \\
& $\nu$ & & $\theta_{\rm maj}$ & $\theta_{\rm min}$ & PA & $\theta_{\rm maj}$ & $\theta_{\rm min}$ & PA & $\theta_{\rm maj}$ & $\theta_{\rm min}$ & PA \\
& (GHz) & & (arcsec) & (arcsec) & (degree) & (arcsec) & (arcsec) & (degree) & (arcsec) & (arcsec) & (degree) \\
(1) & (2) & (3) & (4) & (5) & (6) & (7) & (8) & (9) & (10) & (11) & (12) \\
\hline
\multirow{8}{*}{GMRT} & \multirow{3}{*}{0.32} & Core & \multirow{3}{*}{10.74} & \multirow{3}{*}{7.60} & \multirow{3}{*}{5.8} & 11.83 & 7.93 & 6.4 & $<$5.37 & $<$3.80 & - \\
& & South & & & & 15.81 & 9.53 & 0.1 & 11.64 & 5.67 & 176.9 \\
& & North & & & & 15.60 & 13.00 & 3.9 & 11.30 & 10.50 & 177.0 \\
\cline{2-12}
& \multirow{3}{*}{0.65} & Core & \multirow{3}{*}{4.04} & \multirow{3}{*}{3.78} & \multirow{3}{*}{$-$9.2} & 4.28 & 3.76 & 176.8 & $<$2.02 & $<$1.89 & - \\
& & South & & & & 7.10 & 4.96 & 174.9 & 5.85 & 3.21 & 175.3 \\
& & North & & & & 12.55 & 8.68 & 6.3 & 11.89 & 7.80 & 6.6 \\
\cline{2-12}
& \multirow{2}{*}{1.26} & Core & \multirow{2}{*}{3.40} & \multirow{2}{*}{2.95} & \multirow{2}{*}{28.2} & 3.71 & 3.00 & 18.3 & $<$1.70 & $<$1.48 & - \\
& & South & & & & 5.92 & 4.17 & 134.6 & 5.11 & 2.46 & 132.4 \\
\hline
\multirow{2}{*}{VLASS} & \multirow{2}{*}{3.0} & \multirow{2}{*}{Core} & 3.33 & 2.10 & $-$6.4 & 3.54 & 2.07 & 171.2 & $<$1.67 & $<$1.05 & - \\
& & & 4.03 & 2.07 & $-$6.7 & 4.26 & 2.14 & 172.5 & $<$2.02 & $<$1.04 & - \\
\hline
\multirow{3}{*}{VLA} & \multirow{3}{*}{5.5} & Core & \multirow{3}{*}{4.40} & \multirow{3}{*}{2.07} & \multirow{3}{*}{$-$4.7} & 4.34 & 1.99 & 173.3 & $<$2.20 & $<$1.04 & - \\
& & South & & & & 6.61 & 3.91 & 168.9 & 4.97 & 3.26 & 162.0 \\
& & North & & & & 9.87 & 9.23 & 118.0 & 9.50 & 8.20 & 100.0 \\
\hline
\end{tabular}
\end{table*}

\begin{table*}[ht!]
\caption{The flux densities at 0.32, 0.65, and 1.26~GHz with GMRT, 3.0~GHz with VLASS, and 5.5~GHz with VLA C configuration. Columns:
(1) telescope,
(2) frequency,
(3) component,
(4) total flux density,
(5) peak flux density,
(6) background noise,
(7) spectral slope.}
\label{flux_gmrt}
\centering
\scriptsize
\begin{tabular}{ccccccc}
\hline
\hline
Telescope & $\nu$ & Component & $S_{\rm total}$ & $S_{\rm peak}$ & RMS & Slope \\
& (GHz) & & (mJy) & (mJy\,beam$^{-1}$) & (mJy\,beam$^{-1}$) & \\
(1) & (2) & (3) & (4) & (5) & (6) & (7) \\
\hline
\multirow{8}{*}{GMRT} & \multirow{3}{*}{0.32} & Core & 9.46 $\pm$ 0.40 & 8.23 $\pm$ 0.21 & \multirow{3}{*}{0.089} & \multirow{13}{*}{\makecell{$\alpha_{\rm core} = -0.16$ \\ $\alpha_{\rm south} = -0.72$ \\ $\alpha_{\rm north} = -0.68$}} \\
& & South & 5.10 $\pm$ 0.39 & 2.77 $\pm$ 0.14 & & \\
& & North & 3.99 $\pm$ 0.28 & 1.60 $\pm$ 0.08 & & \\
\cline{2-6}
& \multirow{3}{*}{0.65} & Core & 7.44 $\pm$ 0.21 & 7.06 $\pm$ 0.12 & \multirow{3}{*}{0.035} & \\
& & South & 2.60 $\pm$ 0.29 & 1.12 $\pm$ 0.09 & & \\
& & North & 2.71 $\pm$ 0.24 & 0.38 $\pm$ 0.03 & & \\
\cline{2-6}
& \multirow{2}{*}{1.26} & Core & 5.35 $\pm$ 0.20 & 4.82 $\pm$ 0.11 & \multirow{2}{*}{0.035} & \\
& & South & 1.44 $\pm$ 0.14 & 0.59 $\pm$ 0.04 & & \\
\cline{1-6}
\multirow{2}{*}{VLASS} & \multirow{2}{*}{3.0} & \multirow{2}{*}{Core} & 7.03 $\pm$ 0.24 & 6.70 $\pm$ 0.13 & 0.160 & \\
& & & 6.27 $\pm$ 0.36 & 5.71 $\pm$ 0.18 & 0.150 & \\
\cline{1-6}
\multirow{3}{*}{VLA} & \multirow{3}{*}{5.5} & Core & 5.09 $\pm$ 0.02 & 5.09 $\pm$ 0.02 & \multirow{3}{*}{0.008} & \\
& & South & 0.64 $\pm$ 0.05 & 0.23 $\pm$ 0.01 & & \\
& & North & 0.59 $\pm$ 0.06 & 0.06 $\pm$ 0.01 & & \\
\hline
\end{tabular}
\end{table*}

\section{Results} \label{result_section}

\subsection{Radio morphology}

The VLBA maps (Figure~\ref{map_vlba}) show that the object has extended emission in the southeast direction on mas scales.
The 1.5\,GHz map includes three components. C1 and C2 are unresolved with a peak/total flux ratio of 1. C3 is an extended component with a peak/total flux ratio of 0.7.
At 5\,GHz, C1 and C2 are detected and marginally resolved, and C3 is probably resolved out.
C1 is the compact core emission with a peak/total flux ratio of 0.9, and C2 is more extended with a peak/total flux ratio of 0.7.
The 8.5\,GHz map shows only the C1 component, while C2 is likely resolved out on the 8.5\,GHz resolution.
There is still extended emission detected with a peak/total flux ratio of 0.6, though it can not be resolved as a separate component.

The VLBA 5 and 8.5 GHz core component C1 almost overlap (Figure~\ref{overlap}).
The offsets of their VLBA coordinates from the {\it Gaia} position \citep{Gaia2016,Gaia2023} are about 0.8--0.9~mas (see Table~\ref{size_vlba}), which are consistent with the {\it Gaia} astrometric uncertainty ($\sim$ 1--10~mas) \citep{Khamitov2022}.
The extended components C2 at 5.5~GHz and C3 at 1.5~GHz are about 3 and 15 mas away from the {\it Gaia} position respectively, which suggests that the extended emission is a structure extending from the central core to the southeast direction at a distance of $\sim$ 15~mas.

The GMRT maps (Figure~\ref{map_gmrt}) show a very large jet structure, about 200~kpc, in the south-north direction, which is consistent with our earlier VLA observations \citep{Chen2020}.
The core component is very compact at 0.32--5.5~GHz, with a peak/total flux ratio of 0.9--1.
The two VLASS observations give fluxes of 7.03~mJy in 2019 and 6.27~mJy in 2022, which are consistent with the VLASS calibration uncertainty of $\sim$ 20\% \citep{Lacy2020}.

The southern lobe is likely produced by the approaching jet and is detected at 0.32, 0.65, and 1.26~GHz with the GMRT and at 5.5~GHz with the VLA C configuration.
It shows an edge-brightened FRII-like morphology in the GMRT and VLA images.
The emission is extended, with a peak/total flux ratio of 0.4--0.5.
The northern lobe, which is likely produced by the receding jet, is more diffuse, with a peak/total flux ratio of 0.1--0.4.
It is only detected at 0.32 and 0.65~GHz with the GMRT and 5.5~GHz with the VLA C configuration, and is not edge-brightened.
At 1.26~GHz, if we assume a slope of $-0.7$ (see Table~\ref{flux_gmrt}), the derived peak flux density is higher than the 5$\sigma$ detection criterion.
Thus the non-detection is probably because the emission is resolved out in the GMRT Band 5 after tapering to a resolution of $\sim$ 3~arcsec.
The northern lobe is still undetected when tapering to a lower resolution, which is because of a higher noise level leading to the derived peak flux density below the 5$\sigma$ criterion.
Both lobes are not detected with the VLASS due to the emission being below the VLASS detection limit and/or being resolved out on the VLASS resolution.

\subsection{Radio spectra}

\begin{figure}
\includegraphics[width=\columnwidth, trim={7cm, 0cm, 8cm, 2cm}, clip]{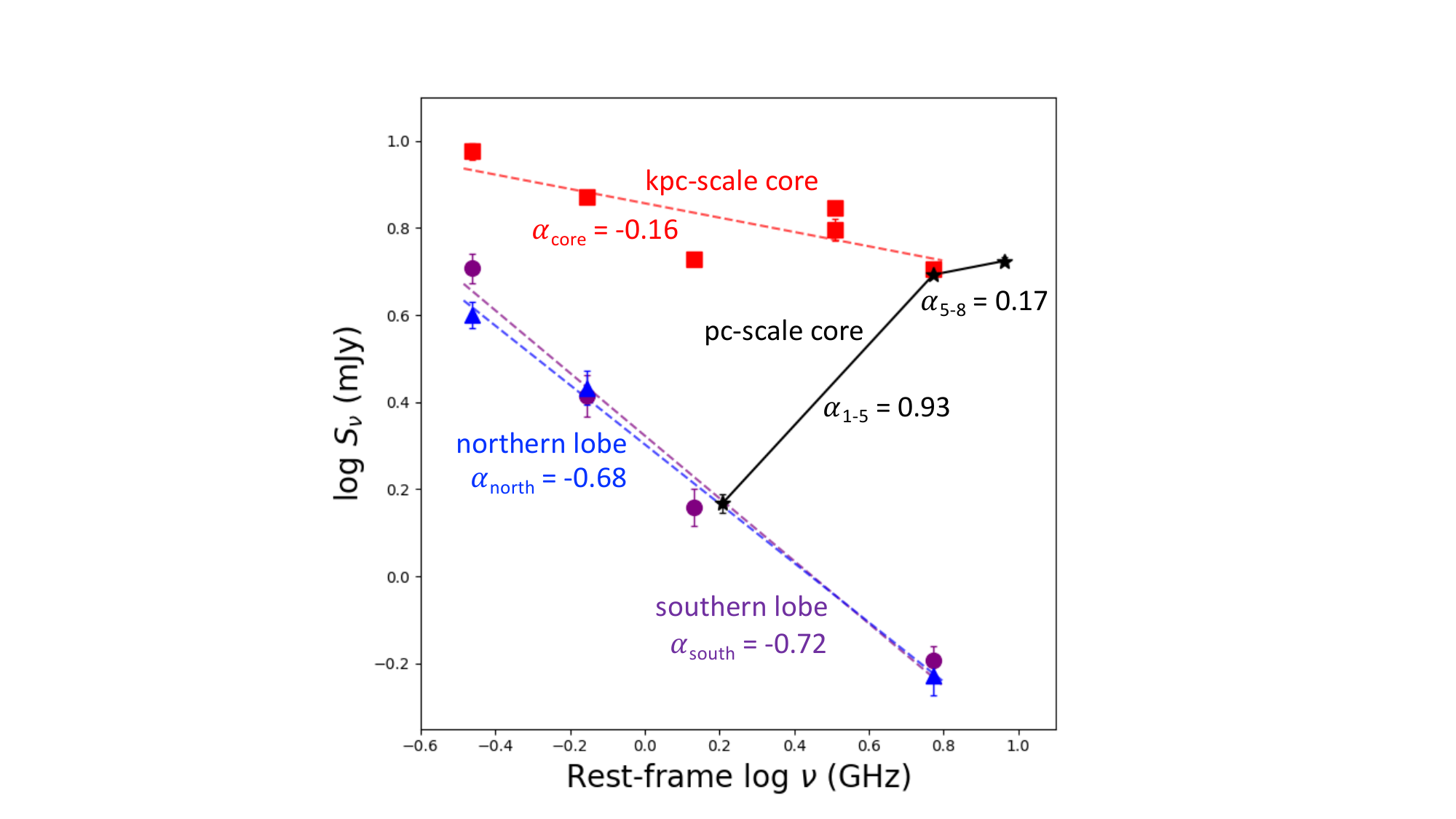}
\caption{The radio spectrum combining observations with VLBA at 1.5--8.5~GHz, GMRT at 0.32--1.26~GHz, VLASS at 3.0~GHz, and VLA C configuration at 5.5~GHz. 
The black stars and solid lines represent the VLBA observations, which is the pc-scale core spectrum.
The red squares show the core flux from the GMRT, VLASS, and VLA observations, and the red dashed line represents a power-law fit, which is the kpc-scale core spectrum.
The purple circles and blue triangles show the southern and northern lobes from the GMRT and VLA observations, and the purple and blue dashed lines represent a power-law fit.
The spectral slope of each component is labeled.}
\label{spectrum}
\end{figure}

Figure~\ref{spectrum} shows the VLBA spectrum at 1.5--8.5~GHz of the object.
The spectral slopes are measured based on the total flux densities in the tapered maps, which have comparable resolutions of $\sim 5$~mas and cover emissions on similar scales in all three bands.
The slope at 1.5--5.5~GHz is $\alpha_{1-5} = 0.93 \pm 0.04$, and the slope at 5.5--8.5~GHz is $\alpha_{5-8} = 0.17 \pm 0.05$.
Only a single component including C1 and C2 is detected after tapering in all three bands, thus the spectrum of C3 is unknown.

The kpc-scale spectra of the core, southern and northern lobes are also presented in Figure~\ref{spectrum} combining the GMRT, VLASS, and VLA observations.
We modeled the spectra with a single power-law.
The kpc-scale core emission at 0.32--5.5~GHz is rather flat, with a spectral slope $\alpha_{\rm core} = -0.16$.
The southern and northern lobes have spectral slopes of $\alpha_{\rm south} = -0.72$ and $\alpha_{\rm north} = -0.68$, respectively, which are consistent with optically thin synchrotron emission from the jet activity.
Both lobes extend to a distance of $\sim$ 100~kpc, which is completely outside the host galaxy.
Thus the radio emission may be produced by the jet interacting with the surrounding environment, e.g., the intracluster medium (ICM) if the NLS1 is in a galaxy cluster.

\section{Discussion} \label{discuss_section}

\subsection{A pc-scale jet or an outflow?}

The extended emission detected in the VLBA observations could be a jet or an outflow.
In a recent VLBA study of a representative sample of RQ PG quasars \citep{Chen2023}, we interpret the optically thin extended emission with an offset of 10--25~mas from the {\it Gaia} position as outflows, instead of jets, due to the low brightness temperature ($10^6-10^7$~K).
We estimate the brightness temperature of different components by adopting the equation \citep[e.g.][]{Ulvestad2005}
\begin{equation}
T_{\rm B} = 1.8 \times 10^{9} (1+z) \frac{S_\nu}{\nu^2 \theta_{\rm max} \theta_{\rm min}} \, \rm{K} \, ,
\end{equation}
where $S_\nu$ is the flux density in mJy at the observing frequency $\nu$ in GHz, and $\theta_{\rm max}$ and $\theta_{\rm min}$ are the major and minor axes of the source size in mas.
The VLBA core component C1 has $\log T_{\rm B} = 7.9-8.2$~(K) at 5 and 8.5~GHz, while the extended components C2 at 5~GHz and C3 at 1.5~GHz have $\log T_{\rm B} = 7.3-7.4$~(K).
The brightness temperature of the extended components is higher than that of the outflow emission \citep[$T_{\rm B} \sim 10^6-10^7$~K;][]{Chen2023}, and slightly lower than that of the core components.
This suggests that the extended emission originates from a mas-scale jet.
Further, the comparatively high $T_{\rm B}$ ($> 10^7$~K) is in agreement with the one-sided jet interpretation, which is caused by the Doppler boosting and dimming, in the VLBA maps (Figure~\ref{map_vlba}).

In addition, Figure~\ref{overlap} may show the core shift effect observed in AGN jets, which is an effect of a frequency dependent shift of the VLBA core position.
This behavior is predicted by the \citet{Blandford1979} model of a purely synchrotron self-absorbed conical jet in equipartition.
The core position shifts as a function of frequency and is consistent with the $r_{\rm core}(\nu) \propto \nu^{-1/k}$ pattern \citep{Sokolovsky2011,Chamani2023}.
In our VLBA observations, the C1 component at 5.5 and 8.5 GHz is core-dominated.
The radio position is very close to the {\it Gaia} position, with an offset of only 0.8--0.9~mas which is negligible \citep{Porcas2009}.
The C1 and C2 components at 1.5~GHz shift to the south direction, which is the direction of the approaching jet, and offset from the {\it Gaia} position about 6.5~mas.
The core shift is seen between 1.5 and 5.5~GHz, and is not seen between 5.5 and 8.5~GHz, which is consistent with strong self-absorption at 1.5--5.5~GHz and weak self-absorption at 5.5--8.5~GHz.
This phenomenon may support the interpretation of the 1.5~GHz pc-scale core emission as a continuous jet.
Another explanation is a resolution effect where the extended emission to the south of the C1 and C2 components is unresolved at 1.5~GHz but resolved out at 5.5 and 8.5~GHz, which can also result in the morphology seen in Figure~\ref{overlap}.

\subsection{The pc-scale and kpc-scale radio spectra}

The mas-scale core emission shows a rising spectrum at 1.5--8.5~GHz, which indicates an optically thick self-absorbed synchrotron source.
The synchrotron peak is very close to 8.5~GHz given the rather flat 5.5--8.5~GHz slope of 0.2.
Assuming that the magnetic energy density is in equipartition with the AGN photon energy density, we can constrain the size of the radio core emission at 8.5~GHz.
The implied size of an optically thick synchrotron source is \citep[eq.22 in][]{Laor2008}
\begin{equation}
R = 0.47 L_{30}^{0.4} L_{46}^{0.1} \nu_{\rm p}^{-1} \, ,
\label{rs}
\end{equation}
where $R$ is the radius of the radio-sphere in pc, $\nu_{\rm p} = 8.5$ is the self-absorbed frequency in GHz, $L_{30} = 0.75$ is the radio luminosity density at 8.5~GHz in units of 10$^{30}$\,erg\,s$^{-1}$\,Hz$^{-1}$ from our VLBA observations, and $L_{46} = 0.07$ is the bolometric luminosity in units of 10$^{46}$\,erg\,s$^{-1}$ from \citet{Chen2018}.
We got the physical size of the optically thick radio core of $\sim$ 0.04\,pc or 48 light days.
The size of the BLR is $\sim$ 0.03\,pc or 30 light days, given $\lambda L_{\lambda}$(5100\AA) $\sim 10^{44}$~erg\,s$^{-1}$ \citep{Chen2018}, which is estimated based on the relation between the 5100{\AA} continuum luminosity and the BLR radius \citep{Bentz2013}.
Thus the radio core is on the scale of the BLR or smaller.
Such a small size is likely to be associated with the jet base or the accretion disk corona \citep{Chen2023}, which may in fact physically coincide \citep{Blundell1998,Merloni2002,King2017}.

The kpc-scale spectrum is a superposition of multiple components.
The VLA core flux and the VLBA total flux at 5.5~GHz are very similar, 5.09~mJy and 5.33~mJy respectively (see Tables \ref{flux_vlba} and \ref{flux_gmrt}), which indicates that the pc-scale emission dominates the kpc-scale core at 5.5~GHz.
However, the pc-scale emission becomes self-absorbed at 1--5~GHz, so there must be another component on kpc scales, which dominates below 5~GHz and contributes to the rather flat ($\alpha_{\rm core} = -0.16$) spectrum.
Such a component probably originates from the jet emission, which is unresolved on scales of $\sim$ 3~arcsec.
This is in agreement with a previous kpc-scale radio spectrum study of NLS1s by \citet{Chen2022a}, which suggests that most RQ NLS1s have a synchrotron component possibly coming from a compact jet/wind with a spectral turnover at $\sim$ 1~GHz.
The jet emission may become optically thin above $\sim$ 1~GHz, and be partly resolved out in our VLBA observations at 1--9~GHz.

We note in passing that, free-free emission may also contribute to the kpc-scale emission, since the spectral slope is consistent with the free-free emission from photoionized gas slope of $\alpha = -0.1$.
We estimate the free-free emission at 5~GHz using the relation \citep[eq.22 in][]{Baskin2021}
\begin{equation}
\log \nu L_{\rm 5GHz} = \log L_{\rm [O III]} - 4.5 \, \rm{(erg\,s^{-1})} \, ,
\end{equation}
where $\nu L_{\rm 5GHz}$ is the radio 5~GHz luminosity produced by free-free emission, and $L_{\rm [O III]}$ is the [O\,III]$\lambda$5007 emission line luminosity.
In this object, $\log L_{\rm [O III]}$ = 41.0 (erg\,s$^{-1}$) \citep{Chen2018}, we thus get $\log \nu L_{\rm 5GHz}$ = 36.5 (erg\,s$^{-1}$) produced by free-free emission, which falls short by three orders of magnitude compared to the measured value of $\log \nu L_{\rm 5GHz}$ = 39.6 (erg\,s$^{-1}$) \citep{Chen2020}.
Generally, free-free emission is more likely to become detectable in mm wavebands \citep{Baskin2021}.

\subsection{The pc-scale and kpc-scale jet misalignment}

The pc-scale jet is in the southeast direction with an angle of 160$^{\circ}$ from the north in counter-clockwise, while the kpc-scale jet lies generally in the south-north direction and the southern jet has an angle of 190$^{\circ}$ from the north. There are three possible scenarios to explain the mismatch of the pc and kpc direction of the southern jet by 30$^{\circ}$, an intermittent jet, a precessing jet, or a deflected jet.

In the first scenario, the jet may come from an intermittent AGN activity. The kpc-scale jet may be the relic emission of past launching activity, and the pc-scale jet may have started recently.
The jets launched previously and recently can have different directions.
This scenario has been invoked to explain the morphology of different kinds of radio sources \citep[e.g.][]{Schoenmakers2000,Kaiser2000,Kharb2006,Murgia2011,Orru2015,Congiu2020}.
The physical processes which produce this change in the AGN jet activity are still debated.
Possible explanations are galaxy merger, in which the infalling gas could cause instability of the accretion flow and quenching and restarting the jet activity \citep{Schoenmakers2000,Nandi2017}, or an internal instability of the accretion, which can change the orientation of the jet \citep{Hernandez-Garcia2017}.
However, the kpc-scale lobes have $\alpha = -0.7$ at 0.32--5.5~GHz, which indicates that synchrotron cooling in this frequency range does not dominate, or the lobes are fed by intermittent fuelling \citep{Sridhar2020,Silpa2021}.
We can assume a simplified picture that the magnetic energy in an isotropic radio lobe equals to the jet injection energy.
Based on the magnetic field derived from such a simple assumption, the synchrotron cooling time is longer than the age of the Universe.
Thus it agrees with the observed slopes that the aging of the relativistic electron population does not dominate.

In the second scenario, the jet axis is in a continuous precession.
Such a phenomenon produces peculiar extended radio structures, such as S or X shapes, which are commonly observed in Seyfert galaxies and also in radio galaxies and blazars \citep{Kukula1995,Nagar1999,Thean2000,Kharb2006,Kharb2010b,Kharb2010a}.
Possible reasons for the jet precession are a binary supermassive BH \citep{Roos1988,Nandi2021}, or a warped disk \citep{Pringle1996,Pringle1997}.
However, there is currently no information about the emission between the pc and 100~kpc scales.
Future radio observations on sub-arcsec scales, such as e-MERLIN and VLA A configuration, are necessary to explore whether and how the pc-scale and kpc-scale jets connect to each other.

The third scenario is an external effect that the jet is deflected by an interaction with the surrounding environment, e.g., a molecular cloud in the interstellar medium (ISM).
This phenomenon is seen in some radio galaxies \citep{VanBreugel1985,Tingay1996,Tingay1997}.
The jet interactions do not halt or disrupt the jet completely, but change the direction by up to a few tens of degrees, through a jet deflection.
The kpc-scale jet in the object is close to symmetric.
If a jet deflection happens, the ambient medium around the pc-scale jet is distributed symmetrically, which leads to a nearly symmetrical kpc-scale jet deflected on both sides.

\begin{figure}
\includegraphics[width=\columnwidth, trim={5cm, 0cm, 5cm, 0cm}, clip]{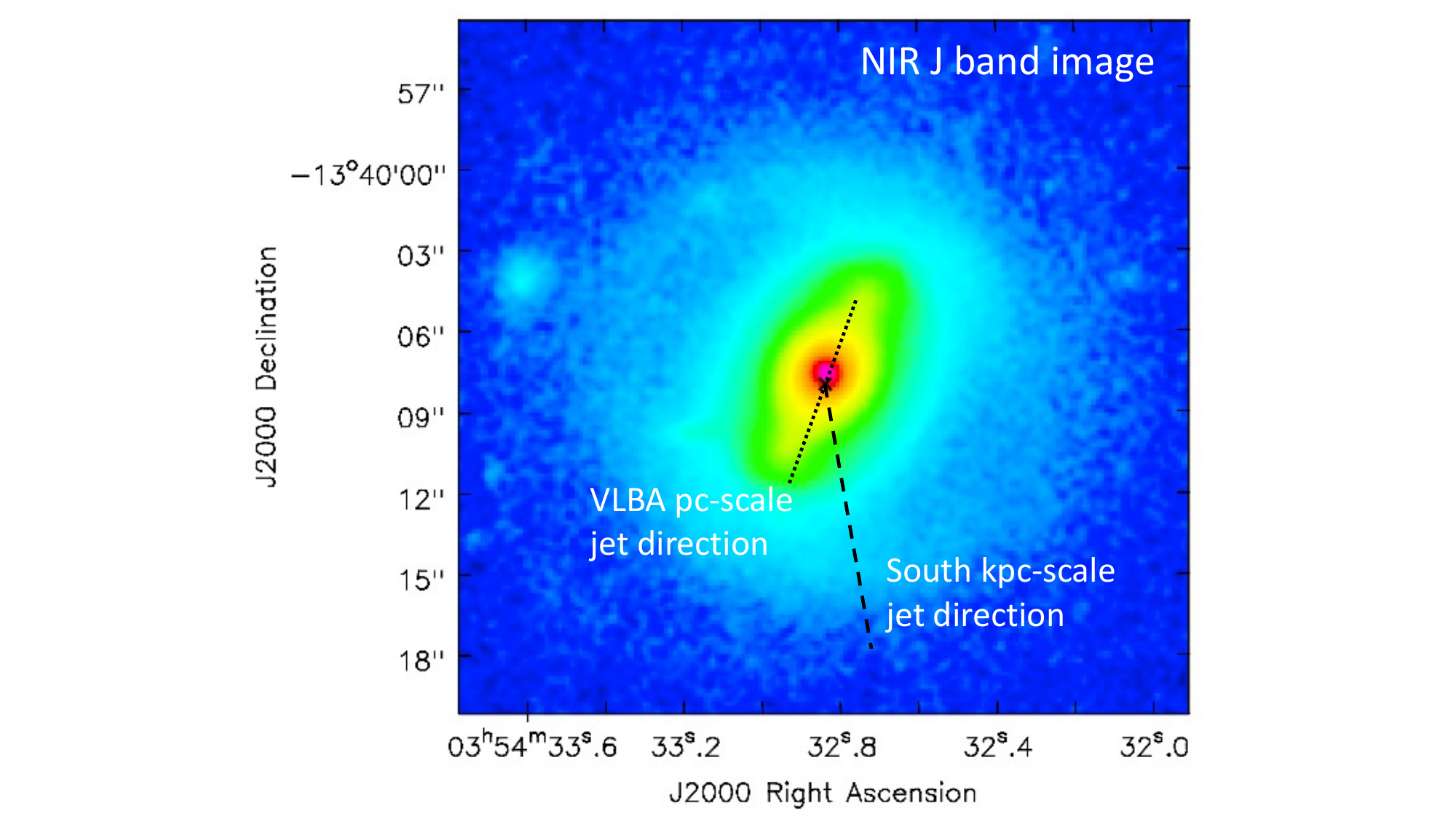}
\caption{The near-infrared $J$-band image observed with the {\it Magellan Baade} telescope \citep{Vietri2022}. The black cross marks the VLBA and the {\it Gaia} positions, which are $\lesssim$ 1~mas offset. The pc-scale and kpc-scale jet directions are marked in black dotted line and black dashed line, respectively, for comparison. The pc-scale jet direction is roughly aligned with the host galaxy bar with only a few degrees offset. We note that there is a small position offset ($\sim$ one third arcsec) between the {\it Gaia} and the $J$-band image centers, which is likely an artefact.}
\label{Magellan}
\end{figure}

We further compare the orientation of the host galaxy and the pc-scale and kpc-scale jets.
Figure~\ref{Magellan} shows the near-infrared $J$-band image observed with the {\it Magellan Baade} telescope \citep{Vietri2022}, and the pc-scale and kpc-scale jet directions.
J0354$-$1340 is hosted in a barred spiral galaxy as found in a host galaxy photometric decomposition study \citep{Vietri2022}.
It is interesting that the pc-scale jet direction is roughly aligned with the host galaxy bar with only a few degrees offset.
Thus it is possible that the pc-scale jet interacts with the symmetrically distributed ISM in the bar, which may support the deflected jet explanation for the misalignment of the pc-scale and kpc-scale jets.
Future integral field spectroscopy observations, which map the distribution of the surrounding gas, will help to explore this scenario.

\subsection{How rare are jetted RQ or RI AGN?}

J0354$-$1340 shows a spectacular 100-kpc two-sided jet, which is commonly present in RL AGN, but is rare in RQ or RI AGN.
Is the detection of the jet structure possible just because it is nearby and thus bright, and thus our instruments are able to detect and resolve it?
In our previous VLA C configuration survey of a sample of 49 NLS1s detected at 5.5~GHz \citep{Chen2020}, the average sensitivity is about 10~$\mu$Jy\,beam$^{-1}$.
The flux density of the lobes of this object is about 0.5~mJy, which is a factor of 10 fainter than that of the core of about 5~mJy.
Let us assume that all the detected NLS1s have lobes that are a factor of 10 fainter than the core.
If we want to detect the lobes at a 5$\sigma$ level in our previous VLA survey, the lobes need to be brighter than 0.05~mJy and the core needs to be brighter than 0.5~mJy.
The flux densities in our previous VLA survey range from 0.05 to 21.28 mJy, and about half of the objects (24/49) have a flux density greater than 0.5~mJy. We found only one such object out of the 24 objects with a flux density higher than 0.5~mJy, which indicates that such large-jet NLS1s occur in only $\sim$ 4\% (1/24), or even less, of the NLS1 population.

\begin{figure}
\includegraphics[width=\columnwidth, trim={0cm, 0cm, 1cm, 1cm}, clip]{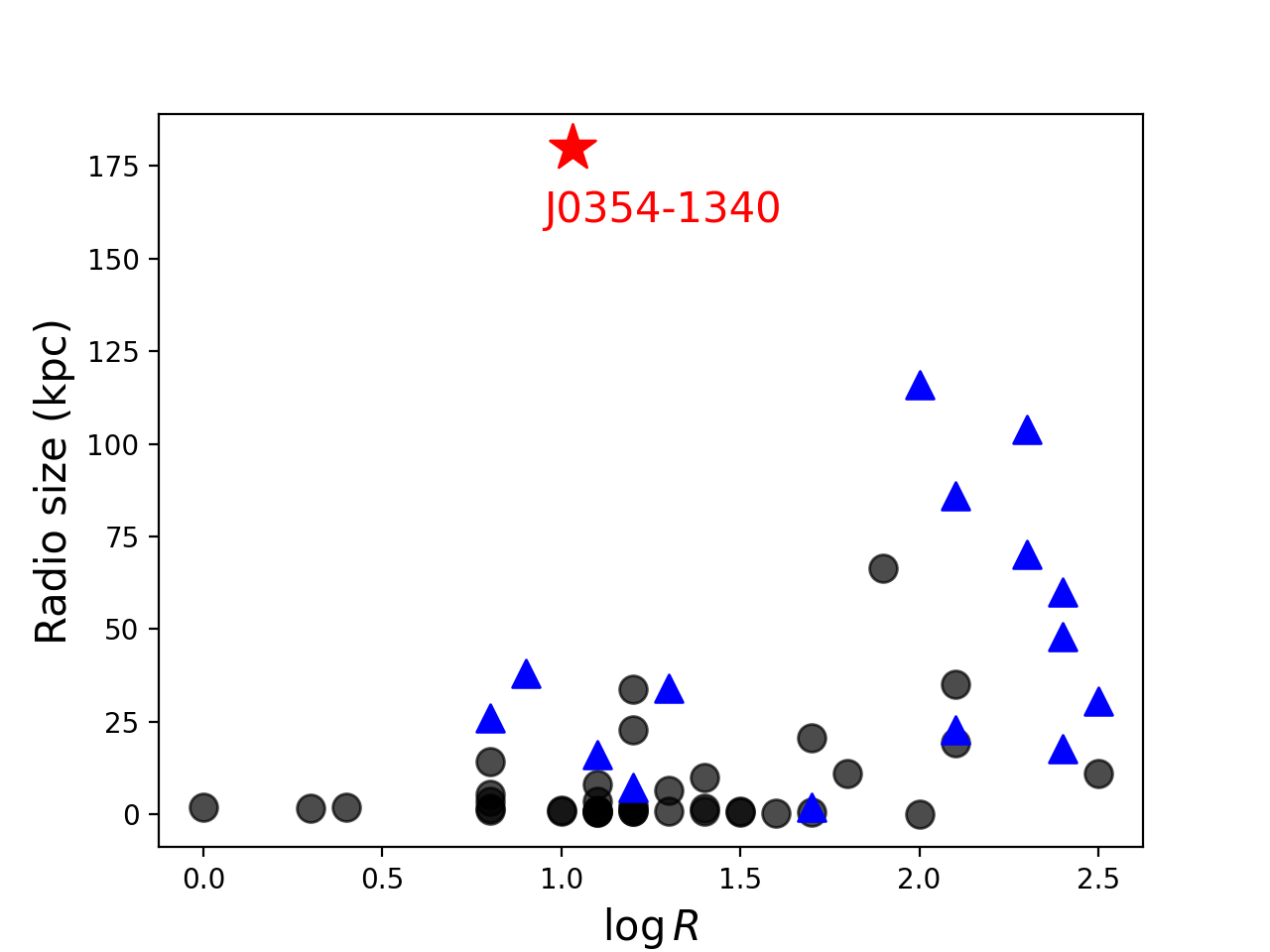}
\caption{A comparison of the radio source size and the radio loudness $R$ in different objects. The red star represents J0354$-$1340. The black circles represent quasars from \citet{Jarvis2021}, which includes a sample of RL and RQ quasars at $z < 0.2$. The blue triangles represent NLS1s from \citet[][Table~1 therein]{Rakshit2018} and \citet[][Table~A1 therein]{Singh2018}, which list NLS1s with kpc-scale structures. It is evident that the large jet structure in J0354$-$1340 is exceptional compared to the other quasars and NLS1s in the RQ and RI regimes.}
\label{jet_size}
\end{figure}

Such large-scale jets are apparently rare in AGN with $R \lesssim 1$, but are they more common in objects with $1 < R \lesssim 10$?
Figure~\ref{jet_size} shows a comparison of the radio source size and the radio loudness $R$ in different objects.
The quasars are from \citet{Jarvis2021}, which includes a sample of RL and RQ quasars at $z < 0.2$.
The NLS1s are from \citet[][Table~1 therein]{Rakshit2018} and \citet[][Table~A1 therein]{Singh2018}, which list NLS1s with kpc-scale structures.
The radio size measured in these objects are based on observations with a resolution of a few arcsec, e.g.\ the Faint Images of the Radio Sky at Twenty-Centimeters \citep[FIRST,][]{Helfand2015}.
We re-calculated the $R$ value in a uniform way.
The 5~GHz flux is converted from the NVSS 1.4~GHz flux assuming $\alpha = -0.7$.
The B-band flux is converted from the Sloan Digital Sky Survey (SDSS) $g$ and $r$ magnitudes \citep{Jester2005}.
We exclude the objects with an upper limit size and include the objects with $\log R < 3$.
It is evident that the object with a large radio size tends to be RL with a high $R$ value, and the large jet in J0354$-$1340 is exceptional compared to the other quasars and NLS1s with $0 < \log R < 3$.

A recent study of a representative sample of RQ PG quasars found a few transition objects between RL and RQ at $\log M_{\rm BH}/M_{\odot} \gtrsim 8.5$ and $\log L_{\rm R}/L_{\rm X} \gtrsim -4$ \citep[see Figure 4 in][]{Chen2023}.
In terms of radio properties, J0354$-$1340 is similar to the PG quasar PG1351+640, which has $R$ = 4.3 \citep{Kellermann1989} and $\log L_{\rm R}/L_{\rm X} = -3.9$ \citep{Chen2023}.
They both show pc-scale jets with the VLBA.
The core emission has $T_{\rm B} \sim 10^8$~K, and the $T_{\rm B}$ of the jet is comparable to or slightly lower than that of the core.
The differences are that J0354$-$1340 has much lower $M_{\rm BH}$ and higher $L/L_{\rm Edd}$ than PG1351+640, $M_{\rm BH} = 10^{7.0} M_{\odot}$ and $L/L_{\rm Edd} = 0.60$ for J0354$-$1340, and $M_{\rm BH} = 10^{8.5} M_{\odot}$ and $L/L_{\rm Edd} = 0.05$ for PG1351+640.
Additionally, the jet in PG1351+640 moves at a mildly relativistic speed of $\sim 0.5c$, as indicated in a recent proper motion study by \citet{Wang2023b}.
Does the pc-scale jet in J0354$-$1340 also move at a sub-relativistic speed?
Is it generally true that jets in RQ or RI AGN, if present, move at a sub-relativistic speed?
Future proper motion monitoring on time scales of years may answer these questions.

\section{Summary} \label{summary_section}

In this work, we present new VLBA 1.5--8.5~GHz and GMRT 0.32--1.26~GHz observations of J0354$-$1340, which is a RQ or RI NLS1 with a 100-kpc two-sided radio jet and is hosted in a barred spiral galaxy. In combination with archival VLASS observations at 3.0~GHz and VLA C configuration observations at 5.5~GHz, we study the jet emission and the core spectrum on pc and kpc scales. The main results are summarized as follows.

1. The extended emission observed with the VLBA in the southeast direction of the core likely originates from a pc-scale jet.
The kpc-scale jet observed with the VLA and the GMRT is in the south-north direction.
The misalignment between the pc-scale and kpc-scale jets may be caused by an intermittent jet, a precessing jet, or a deflected jet.

2. The pc-scale spectrum observed with the VLBA suggests an optically thick synchrotron source of $\sim$ 0.04~pc, which can be naturally explained by the jet base or the corona.
The pc-scale emission dominates the kpc-scale core emission above $\sim$ 5~GHz, but it becomes self-absorbed and negligible below $\sim$ 5~GHz.
A sub-kpc scale jet, which is unresolved with the VLA and the GMRT, probably dominates the kpc-scale core emission below $\sim$ 5~GHz.

Future radio observations on sub-arcsec scales, such as e-MERLIN and VLA A configuration, can reveal whether and how the pc-scale and kpc-scale jets connect to each other.
Moreover, proper motion monitoring on time scales of years can help to investigate whether the jet moves at a lower speed in RQ or RI AGN compared to RL AGN.
It is also worth exploring how common such large-scale jets are in the boundary between RQ and RL AGN.

\begin{acknowledgments}

We thank the anonymous referee for a thorough review and helpful suggestions leading to the improvement of this work.
S.C. is supported in part by a Technion fellowship.
A.L. acknowledges support by the Israel Science Foundation (grant no.1008/18).
E.B. acknowledges support by a Center of Excellence of the Israel Science Foundation (grant no.2752/19).
P.K. and S.S. acknowledge the support of the Department of Atomic Energy, Government of India, under the project 12-R\&D-TFR-5.02-0700.
S.S. acknowledges financial support from Millenium Nucleus NCN19\_058 (TITANs).
M.B. and L.C. acknowledge the support by the ESO science support discretionary fund (SSDF) 2023.
MFG is supported by the Shanghai Pilot Program for Basic Research-Chinese Academy of Science, Shanghai Branch (JCYJ-SHFY-2021-013), the National SKA Program of China (Grant No.\,2022SKA0120102), the science research grants from the China Manned Space Project with No.\,CMSCSST-2021-A06, and the Original Innovation Program of the Chinese Academy of Sciences (E085021002).
MHZ is supported by the Natural Science Foundation of Jiangxi (No.\,20232BAB211024), and the Doctoral Scientific Research Foundation of Shangrao Normal University (Grant No.\,K6000449).
The National Radio Astronomy Observatory is a facility of the National Science Foundation operated under cooperative agreement by Associated Universities, Inc.
We thank the staff of the GMRT that made these observations possible. GMRT is run by the National Centre for Radio Astrophysics of the Tata Institute of Fundamental Research.
This work has made use of data from the European Space Agency (ESA) mission {\it Gaia} (\url{https://www.cosmos.esa.int/gaia}), processed by the {\it Gaia} Data Processing and Analysis Consortium (DPAC, \url{https://www.cosmos.esa.int/web/gaia/dpac/consortium}). Funding for the DPAC has been provided by national institutions, in particular the institutions participating in the {\it Gaia} Multilateral Agreement.
This research has made use of data obtained from XMMSL2, the Second XMM-Newton Slew Survey Catalogue, produced by members of the XMM SOC, the EPIC consortium, and using work carried out in the context of the EXTraS project (``Exploring the X-ray Transient and variable Sky'', funded from the EU's Seventh Framework Programme under grant agreement no.607452).

\end{acknowledgments}


\vspace{5mm}
\facilities{VLBA, GMRT}


\software{AIPS \citep{Greisen2003}, CASA \citep{CASA2022}, astropy \citep{Astropy2022}}

\bibliography{main.bbl}

\begin{thebibliography}{}
\expandafter\ifx\csname natexlab\endcsname\relax\def\natexlab#1{#1}\fi
\providecommand{\url}[1]{\href{#1}{#1}}
\providecommand{\dodoi}[1]{doi:~\href{http://doi.org/#1}{\nolinkurl{#1}}}
\providecommand{\doeprint}[1]{\href{http://ascl.net/#1}{\nolinkurl{http://ascl.net/#1}}}
\providecommand{\doarXiv}[1]{\href{https://arxiv.org/abs/#1}{\nolinkurl{https://arxiv.org/abs/#1}}}

\bibitem[{{Abdo} {et~al.}(2009{\natexlab{a}}){Abdo}, {Ackermann}, {Ajello},
  {Axelsson}, {Baldini}, {Ballet}, {Barbiellini}, {Bastieri}, {Battelino},
  {Baughman}, {Bechtol}, {Bellazzini}, {Bloom}, {Bonamente}, {Borgland},
  {Bregeon}, {Brez}, {Brigida}, {Bruel}, {Caliandro}, {Cameron}, {Caraveo},
  {Casandjian}, {Cavazzuti}, {Cecchi}, {Chekhtman}, {Cheung}, {Chiang},
  {Ciprini}, {Claus}, {Cohen-Tanugi}, {Collmar}, {Conrad}, {Costamante},
  {Dermer}, {de Angelis}, {de Palma}, {Digel}, {Silva}, {Drell}, {Dubois},
  {Dumora}, {Farnier}, {Favuzzi}, {Focke}, {Foschini}, {Frailis}, {Fuhrmann},
  {Fukazawa}, {Funk}, {Fusco}, {Gargano}, {Gehrels}, {Germani}, {Giebels},
  {Giglietto}, {Giordano}, {Giroletti}, {Glanzman}, {Grenier}, {Grondin},
  {Grove}, {Guillemot}, {Guiriec}, {Hanabata}, {Harding}, {Hartman},
  {Hayashida}, {Hays}, {Hughes}, {J{\'o}hannesson}, {Johnson}, {Johnson},
  {Johnson}, {Kamae}, {Katagiri}, {Kataoka}, {Kerr}, {Kn{\"o}dlseder}, {Kuehn},
  {Kuss}, {Lande}, {Latronico}, {Lemoine-Goumard}, {Longo}, {Loparco}, {Lott},
  {Lovellette}, {Lubrano}, {Madejski}, {Makeev}, {Max-Moerbeck}, {Mazziotta},
  {McConville}, {McEnery}, {Meurer}, {Michelson}, {Mitthumsiri}, {Mizuno},
  {Monte}, {Monzani}, {Morselli}, {Moskalenko}, {Murgia}, {Nolan}, {Norris},
  {Nuss}, {Ohsugi}, {Omodei}, {Orlando}, {Ormes}, {Paneque}, {Panetta},
  {Parent}, {Pavlidou}, {Pearson}, {Pepe}, {Pesce-Rollins}, {Piron}, {Porter},
  {Rain{\`o}}, {Rando}, {Razzano}, {Readhead}, {Reimer}, {Reimer}, {Reposeur},
  {Richards}, {Ritz}, {Rodriguez}, {Romani}, {Ryde}, {Sadrozinski}, {Sambruna},
  {Sanchez}, {Sander}, {Parkinson}, {Scargle}, {Schalk}, {Sgr{\`o}}, {Smith},
  {Spandre}, {Spinelli}, {Starck}, {Stevenson}, {Strickman}, {Suson},
  {Tagliaferri}, {Takahashi}, {Tanaka}, {Thayer}, {Thompson}, {Tibaldo},
  {Tibolla}, {Torres}, {Tosti}, {Tramacere}, {Uchiyama}, {Usher}, {Vilchez},
  {Vitale}, {Waite}, {Winer}, {Wood}, {Ylinen}, {Zensus}, {Ziegler}, {Fermi/LAT
  Collaboration}, {Ghisellini}, {Maraschi}, {Tavecchio}, \&
  {Angelakis}}]{Abdo2009a}
{Abdo}, A.~A., {Ackermann}, M., {Ajello}, M., {et~al.} 2009{\natexlab{a}},
  \apj, 699, 976, \dodoi{10.1088/0004-637X/699/2/976}

\bibitem[{{Abdo} {et~al.}(2009{\natexlab{b}}){Abdo}, {Ackermann}, {Ajello},
  {Axelsson}, {Baldini}, {Ballet}, {Barbiellini}, {Bastieri}, {Baughman},
  {Bechtol}, \& et~al.}]{Abdo2009b}
---. 2009{\natexlab{b}}, \apj, 707, 727, \dodoi{10.1088/0004-637X/707/1/727}

\bibitem[{{Astropy Collaboration} {et~al.}(2022){Astropy Collaboration},
  {Price-Whelan}, {Lim}, {Earl}, {Starkman}, {Bradley}, {Shupe}, {Patil},
  {Corrales}, {Brasseur}, {N{\"o}the}, {Donath}, {Tollerud}, {Morris},
  {Ginsburg}, {Vaher}, {Weaver}, {Tocknell}, {Jamieson}, {van Kerkwijk},
  {Robitaille}, {Merry}, {Bachetti}, {G{\"u}nther}, {Aldcroft},
  {Alvarado-Montes}, {Archibald}, {B{\'o}di}, {Bapat}, {Barentsen},
  {Baz{\'a}n}, {Biswas}, {Boquien}, {Burke}, {Cara}, {Cara}, {Conroy},
  {Conseil}, {Craig}, {Cross}, {Cruz}, {D'Eugenio}, {Dencheva}, {Devillepoix},
  {Dietrich}, {Eigenbrot}, {Erben}, {Ferreira}, {Foreman-Mackey}, {Fox},
  {Freij}, {Garg}, {Geda}, {Glattly}, {Gondhalekar}, {Gordon}, {Grant},
  {Greenfield}, {Groener}, {Guest}, {Gurovich}, {Handberg}, {Hart},
  {Hatfield-Dodds}, {Homeier}, {Hosseinzadeh}, {Jenness}, {Jones}, {Joseph},
  {Kalmbach}, {Karamehmetoglu}, {Ka{\l}uszy{\'n}ski}, {Kelley}, {Kern},
  {Kerzendorf}, {Koch}, {Kulumani}, {Lee}, {Ly}, {Ma}, {MacBride}, {Maljaars},
  {Muna}, {Murphy}, {Norman}, {O'Steen}, {Oman}, {Pacifici}, {Pascual},
  {Pascual-Granado}, {Patil}, {Perren}, {Pickering}, {Rastogi}, {Roulston},
  {Ryan}, {Rykoff}, {Sabater}, {Sakurikar}, {Salgado}, {Sanghi}, {Saunders},
  {Savchenko}, {Schwardt}, {Seifert-Eckert}, {Shih}, {Jain}, {Shukla}, {Sick},
  {Simpson}, {Singanamalla}, {Singer}, {Singhal}, {Sinha}, {Sip{\H{o}}cz},
  {Spitler}, {Stansby}, {Streicher}, {{\v{S}}umak}, {Swinbank}, {Taranu},
  {Tewary}, {Tremblay}, {de Val-Borro}, {Van Kooten}, {Vasovi{\'c}}, {Verma},
  {de Miranda Cardoso}, {Williams}, {Wilson}, {Winkel}, {Wood-Vasey}, {Xue},
  {Yoachim}, {Zhang}, {Zonca}, \& {Astropy Project Contributors}}]{Astropy2022}
{Astropy Collaboration}, {Price-Whelan}, A.~M., {Lim}, P.~L., {et~al.} 2022,
  \apj, 935, 167, \dodoi{10.3847/1538-4357/ac7c74}

\bibitem[{{Bagchi} {et~al.}(2014){Bagchi}, {Vivek}, {Vikram}, {Hota}, {Biju},
  {Sirothia}, {Srianand}, {Gopal-Krishna}, \& {Jacob}}]{Bagchi2014}
{Bagchi}, J., {Vivek}, M., {Vikram}, V., {et~al.} 2014, \apj, 788, 174,
  \dodoi{10.1088/0004-637X/788/2/174}

\bibitem[{{Baskin} \& {Laor}(2021)}]{Baskin2021}
{Baskin}, A., \& {Laor}, A. 2021, \mnras, 508, 680,
  \dodoi{10.1093/mnras/stab2555}

\bibitem[{{Bentz} {et~al.}(2013){Bentz}, {Denney}, {Grier}, {Barth},
  {Peterson}, {Vestergaard}, {Bennert}, {Canalizo}, {De Rosa}, {Filippenko},
  {Gates}, {Greene}, {Li}, {Malkan}, {Pogge}, {Stern}, {Treu}, \&
  {Woo}}]{Bentz2013}
{Bentz}, M.~C., {Denney}, K.~D., {Grier}, C.~J., {et~al.} 2013, \apj, 767, 149,
  \dodoi{10.1088/0004-637X/767/2/149}

\bibitem[{{Berton} {et~al.}(2018){Berton}, {Congiu}, {J{\"a}rvel{\"a}},
  {Antonucci}, {Kharb}, {Lister}, {Tarchi}, {Caccianiga}, {Chen}, {Foschini},
  {L{\"a}hteenm{\"a}ki}, {Richards}, {Ciroi}, {Cracco}, {Frezzato}, {La Mura},
  \& {Rafanelli}}]{Berton2018}
{Berton}, M., {Congiu}, E., {J{\"a}rvel{\"a}}, E., {et~al.} 2018, \aap, 614,
  A87, \dodoi{10.1051/0004-6361/201832612}

\bibitem[{{Best} {et~al.}(2005){Best}, {Kauffmann}, {Heckman}, {Brinchmann},
  {Charlot}, {Ivezi{\'c}}, \& {White}}]{Best2005}
{Best}, P.~N., {Kauffmann}, G., {Heckman}, T.~M., {et~al.} 2005, \mnras, 362,
  25, \dodoi{10.1111/j.1365-2966.2005.09192.x}

\bibitem[{{Blandford} \& {K{\"o}nigl}(1979)}]{Blandford1979}
{Blandford}, R.~D., \& {K{\"o}nigl}, A. 1979, \apj, 232, 34,
  \dodoi{10.1086/157262}

\bibitem[{{Blundell} \& {Beasley}(1998)}]{Blundell1998}
{Blundell}, K.~M., \& {Beasley}, A.~J. 1998, \mnras, 299, 165,
  \dodoi{10.1046/j.1365-8711.1998.01752.x}

\bibitem[{{Boller} {et~al.}(1996){Boller}, {Brandt}, \& {Fink}}]{Boller1996}
{Boller}, T., {Brandt}, W.~N., \& {Fink}, H. 1996, \aap, 305, 53

\bibitem[{{Boller} {et~al.}(2016){Boller}, {Freyberg}, {Tr{\"u}mper}, {Haberl},
  {Voges}, \& {Nandra}}]{Boller2016}
{Boller}, T., {Freyberg}, M.~J., {Tr{\"u}mper}, J., {et~al.} 2016, \aap, 588,
  A103, \dodoi{10.1051/0004-6361/201525648}

\bibitem[{{Boroson}(2005)}]{Boroson2005}
{Boroson}, T. 2005, \aj, 130, 381, \dodoi{10.1086/431722}

\bibitem[{{Boroson} \& {Green}(1992)}]{Boroson1992}
{Boroson}, T.~A., \& {Green}, R.~F. 1992, \apjs, 80, 109,
  \dodoi{10.1086/191661}

\bibitem[{{Caccianiga} {et~al.}(2015){Caccianiga}, {Ant{\'o}n}, {Ballo},
  {Foschini}, {Maccacaro}, {Della Ceca}, {Severgnini}, {March{\~a}}, {Mateos},
  \& {Sani}}]{Caccianiga2015}
{Caccianiga}, A., {Ant{\'o}n}, S., {Ballo}, L., {et~al.} 2015, \mnras, 451,
  1795, \dodoi{10.1093/mnras/stv939}

\bibitem[{{Capetti} {et~al.}(2021){Capetti}, {Laor}, {Baldi}, {Robinson}, \&
  {Marconi}}]{Capetti2021}
{Capetti}, A., {Laor}, A., {Baldi}, R.~D., {Robinson}, A., \& {Marconi}, A.
  2021, \mnras, 502, 5086, \dodoi{10.1093/mnras/stab279}

\bibitem[{{CASA Team} {et~al.}(2022){CASA Team}, {Bean}, {Bhatnagar}, {Castro},
  {Donovan Meyer}, {Emonts}, {Garcia}, {Garwood}, {Golap}, {Gonzalez Villalba},
  {Harris}, {Hayashi}, {Hoskins}, {Hsieh}, {Jagannathan}, {Kawasaki},
  {Keimpema}, {Kettenis}, {Lopez}, {Marvil}, {Masters}, {McNichols},
  {Mehringer}, {Miel}, {Moellenbrock}, {Montesino}, {Nakazato}, {Ott}, {Petry},
  {Pokorny}, {Raba}, {Rau}, {Schiebel}, {Schweighart}, {Sekhar}, {Shimada},
  {Small}, {Steeb}, {Sugimoto}, {Suoranta}, {Tsutsumi}, {van Bemmel},
  {Verkouter}, {Wells}, {Xiong}, {Szomoru}, {Griffith}, {Glendenning}, \&
  {Kern}}]{CASA2022}
{CASA Team}, {Bean}, B., {Bhatnagar}, S., {et~al.} 2022, \pasp, 134, 114501,
  \dodoi{10.1088/1538-3873/ac9642}

\bibitem[{{Cavagnolo} {et~al.}(2010){Cavagnolo}, {McNamara}, {Nulsen},
  {Carilli}, {Jones}, \& {B{\^\i}rzan}}]{Cavagnolo2010}
{Cavagnolo}, K.~W., {McNamara}, B.~R., {Nulsen}, P.~E.~J., {et~al.} 2010, \apj,
  720, 1066, \dodoi{10.1088/0004-637X/720/2/1066}

\bibitem[{{Chamani} {et~al.}(2023){Chamani}, {Savolainen}, {Ros}, {Kovalev},
  {Wiik}, {L{\"a}hteenm{\"a}ki}, {Tornikoski}, \& {Tammi}}]{Chamani2023}
{Chamani}, W., {Savolainen}, T., {Ros}, E., {et~al.} 2023, \aap, 672, A130,
  \dodoi{10.1051/0004-6361/202243435}

\bibitem[{{Chen} {et~al.}(2023){Chen}, {Laor}, {Behar}, {Baldi}, \&
  {Gelfand}}]{Chen2023}
{Chen}, S., {Laor}, A., {Behar}, E., {Baldi}, R.~D., \& {Gelfand}, J.~D. 2023,
  \mnras, 525, 164, \dodoi{10.1093/mnras/stad2289}

\bibitem[{{Chen} {et~al.}(2018){Chen}, {Berton}, {La Mura}, {Congiu}, {Cracco},
  {Foschini}, {Fan}, {Ciroi}, {Rafanelli}, \& {Bastieri}}]{Chen2018}
{Chen}, S., {Berton}, M., {La Mura}, G., {et~al.} 2018, \aap, 615, A167,
  \dodoi{10.1051/0004-6361/201832678}

\bibitem[{{Chen} {et~al.}(2020){Chen}, {J{\"a}rvel{\"a}}, {Crepaldi}, {Zhou},
  {Ciroi}, {Berton}, {Kharb}, {Foschini}, {Gu}, {La Mura}, \&
  {Vietri}}]{Chen2020}
{Chen}, S., {J{\"a}rvel{\"a}}, E., {Crepaldi}, L., {et~al.} 2020, \mnras, 498,
  1278, \dodoi{10.1093/mnras/staa2373}

\bibitem[{{Chen} {et~al.}(2022){Chen}, {Stevens}, {Edwards}, {Laor}, {Gu},
  {Berton}, {J{\"a}rvel{\"a}}, {Kharb}, {Behar}, \& {Su}}]{Chen2022a}
{Chen}, S., {Stevens}, J.~B., {Edwards}, P.~G., {et~al.} 2022, \mnras, 512,
  471, \dodoi{10.1093/mnras/stac530}

\bibitem[{{Collin} \& {Kawaguchi}(2004)}]{Collin2004}
{Collin}, S., \& {Kawaguchi}, T. 2004, \aap, 426, 797,
  \dodoi{10.1051/0004-6361:20040528}

\bibitem[{{Congiu} {et~al.}(2020){Congiu}, {Kharb}, {Tarchi}, {Berton},
  {Caccianiga}, {Chen}, {Crepaldi}, {Di Mille}, {J{\"a}rvel{\"a}}, {Jarvis},
  {La Mura}, \& {Vietri}}]{Congiu2020}
{Congiu}, E., {Kharb}, P., {Tarchi}, A., {et~al.} 2020, \mnras, 499, 3149,
  \dodoi{10.1093/mnras/staa3024}

\bibitem[{{Cracco} {et~al.}(2016){Cracco}, {Ciroi}, {Berton}, {Di Mille},
  {Foschini}, {La Mura}, \& {Rafanelli}}]{Cracco2016}
{Cracco}, V., {Ciroi}, S., {Berton}, M., {et~al.} 2016, \mnras, 462, 1256,
  \dodoi{10.1093/mnras/stw1689}

\bibitem[{{Crenshaw} {et~al.}(2003){Crenshaw}, {Kraemer}, \&
  {Gabel}}]{Crenshaw2003b}
{Crenshaw}, D.~M., {Kraemer}, S.~B., \& {Gabel}, J.~R. 2003, \aj, 126, 1690,
  \dodoi{10.1086/377625}

\bibitem[{{Decarli} {et~al.}(2008){Decarli}, {Dotti}, {Fontana}, \&
  {Haardt}}]{Decarli2008}
{Decarli}, R., {Dotti}, M., {Fontana}, M., \& {Haardt}, F. 2008, \mnras, 386,
  L15, \dodoi{10.1111/j.1745-3933.2008.00451.x}

\bibitem[{{Deo} {et~al.}(2006){Deo}, {Crenshaw}, \& {Kraemer}}]{Deo2006}
{Deo}, R.~P., {Crenshaw}, D.~M., \& {Kraemer}, S.~B. 2006, \aj, 132, 321,
  \dodoi{10.1086/504894}

\bibitem[{{Falcke} {et~al.}(1996){Falcke}, {Sherwood}, \&
  {Patnaik}}]{Falcke1996}
{Falcke}, H., {Sherwood}, W., \& {Patnaik}, A.~R. 1996, \apj, 471, 106,
  \dodoi{10.1086/177956}

\bibitem[{{Foschini}(2014)}]{Foschini2014}
{Foschini}, L. 2014, in International Journal of Modern Physics Conference
  Series, Vol.~28, International Journal of Modern Physics Conference Series,
  1460188, \dodoi{10.1142/S2010194514601884}

\bibitem[{{Foschini} {et~al.}(2015){Foschini}, {Berton}, {Caccianiga}, {Ciroi},
  {Cracco}, {Peterson}, {Angelakis}, {Braito}, {Fuhrmann}, {Gallo}, {Grupe},
  {J{\"a}rvel{\"a}}, {Kaufmann}, {Komossa}, {Kovalev}, {L{\"a}hteenm{\"a}ki},
  {Lisakov}, {Lister}, {Mathur}, {Richards}, {Romano}, {Sievers},
  {Tagliaferri}, {Tammi}, {Tibolla}, {Tornikoski}, {Vercellone}, {La Mura},
  {Maraschi}, \& {Rafanelli}}]{Foschini2015}
{Foschini}, L., {Berton}, M., {Caccianiga}, A., {et~al.} 2015, \aap, 575, A13,
  \dodoi{10.1051/0004-6361/201424972}

\bibitem[{{Gaia Collaboration} {et~al.}(2016){Gaia Collaboration}, {Prusti},
  {de Bruijne}, {Brown}, {Vallenari}, {Babusiaux}, {Bailer-Jones}, {Bastian},
  {Biermann}, {Evans}, {Eyer}, {Jansen}, {Jordi}, {Klioner}, {Lammers},
  {Lindegren}, {Luri}, {Mignard}, {Milligan}, {Panem}, {Poinsignon},
  {Pourbaix}, {Randich}, {Sarri}, {Sartoretti}, {Siddiqui}, {Soubiran},
  {Valette}, {van Leeuwen}, {Walton}, {Aerts}, {Arenou}, {Cropper}, {Drimmel},
  {H{\o}g}, {Katz}, {Lattanzi}, {O'Mullane}, {Grebel}, {Holland}, {Huc},
  {Passot}, {Bramante}, {Cacciari}, {Casta{\~n}eda}, {Chaoul}, {Cheek}, {De
  Angeli}, {Fabricius}, {Guerra}, {Hern{\'a}ndez}, {Jean-Antoine-Piccolo},
  {Masana}, {Messineo}, {Mowlavi}, {Nienartowicz}, {Ord{\'o}{\~n}ez-Blanco},
  {Panuzzo}, {Portell}, {Richards}, {Riello}, {Seabroke}, {Tanga},
  {Th{\'e}venin}, {Torra}, {Els}, {Gracia-Abril}, {Comoretto},
  {Garcia-Reinaldos}, {Lock}, {Mercier}, {Altmann}, {Andrae}, {Astraatmadja},
  {Bellas-Velidis}, {Benson}, {Berthier}, {Blomme}, {Busso}, {Carry},
  {Cellino}, {Clementini}, {Cowell}, {Creevey}, {Cuypers}, {Davidson}, {De
  Ridder}, {de Torres}, {Delchambre}, {Dell'Oro}, {Ducourant}, {Fr{\'e}mat},
  {Garc{\'\i}a-Torres}, {Gosset}, {Halbwachs}, {Hambly}, {Harrison}, {Hauser},
  {Hestroffer}, {Hodgkin}, {Huckle}, {Hutton}, {Jasniewicz}, {Jordan},
  {Kontizas}, {Korn}, {Lanzafame}, {Manteiga}, {Moitinho}, {Muinonen},
  {Osinde}, {Pancino}, {Pauwels}, {Petit}, {Recio-Blanco}, {Robin}, {Sarro},
  {Siopis}, {Smith}, {Smith}, {Sozzetti}, {Thuillot}, {van Reeven}, {Viala},
  {Abbas}, {Abreu Aramburu}, {Accart}, {Aguado}, {Allan}, {Allasia},
  {Altavilla}, {{\'A}lvarez}, {Alves}, {Anderson}, {Andrei}, {Anglada Varela},
  {Antiche}, {Antoja}, {Ant{\'o}n}, {Arcay}, {Atzei}, {Ayache}, {Bach},
  {Baker}, {Balaguer-N{\'u}{\~n}ez}, {Barache}, {Barata}, {Barbier}, {Barblan},
  {Baroni}, {Barrado y Navascu{\'e}s}, {Barros}, {Barstow}, {Becciani},
  {Bellazzini}, {Bellei}, {Bello Garc{\'\i}a}, {Belokurov}, {Bendjoya},
  {Berihuete}, {Bianchi}, {Bienaym{\'e}}, {Billebaud}, {Blagorodnova},
  {Blanco-Cuaresma}, {Boch}, {Bombrun}, {Borrachero}, {Bouquillon}, {Bourda},
  {Bouy}, {Bragaglia}, {Breddels}, {Brouillet}, {Br{\"u}semeister},
  {Bucciarelli}, {Budnik}, {Burgess}, {Burgon}, {Burlacu}, {Busonero}, {Buzzi},
  {Caffau}, {Cambras}, {Campbell}, {Cancelliere}, {Cantat-Gaudin}, {Carlucci},
  {Carrasco}, {Castellani}, {Charlot}, {Charnas}, {Charvet}, {Chassat},
  {Chiavassa}, {Clotet}, {Cocozza}, {Collins}, {Collins}, {Costigan}, {Crifo},
  {Cross}, {Crosta}, {Crowley}, {Dafonte}, {Damerdji}, {Dapergolas}, {David},
  {David}, {De Cat}, {de Felice}, {de Laverny}, {De Luise}, {De March}, {de
  Martino}, {de Souza}, {Debosscher}, {del Pozo}, {Delbo}, {Delgado},
  {Delgado}, {di Marco}, {Di Matteo}, {Diakite}, {Distefano}, {Dolding}, {Dos
  Anjos}, {Drazinos}, {Dur{\'a}n}, {Dzigan}, {Ecale}, {Edvardsson}, {Enke},
  {Erdmann}, {Escolar}, {Espina}, {Evans}, {Eynard Bontemps}, {Fabre},
  {Fabrizio}, {Faigler}, {Falc{\~a}o}, {Farr{\`a}s Casas}, {Faye}, {Federici},
  {Fedorets}, {Fern{\'a}ndez-Hern{\'a}ndez}, {Fernique}, {Fienga}, {Figueras},
  {Filippi}, {Findeisen}, {Fonti}, {Fouesneau}, {Fraile}, {Fraser}, {Fuchs},
  {Furnell}, {Gai}, {Galleti}, {Galluccio}, {Garabato}, {Garc{\'\i}a-Sedano},
  {Gar{\'e}}, {Garofalo}, {Garralda}, {Gavras}, {Gerssen}, {Geyer}, {Gilmore},
  {Girona}, {Giuffrida}, {Gomes}, {Gonz{\'a}lez-Marcos},
  {Gonz{\'a}lez-N{\'u}{\~n}ez}, {Gonz{\'a}lez-Vidal}, {Granvik}, {Guerrier},
  {Guillout}, {Guiraud}, {G{\'u}rpide}, {Guti{\'e}rrez-S{\'a}nchez}, {Guy},
  {Haigron}, {Hatzidimitriou}, {Haywood}, {Heiter}, {Helmi}, {Hobbs},
  {Hofmann}, {Holl}, {Holland}, {Hunt}, {Hypki}, {Icardi}, {Irwin}, {Jevardat
  de Fombelle}, {Jofr{\'e}}, {Jonker}, {Jorissen}, {Julbe}, {Karampelas},
  {Kochoska}, {Kohley}, {Kolenberg}, {Kontizas}, {Koposov}, {Kordopatis},
  {Koubsky}, {Kowalczyk}, {Krone-Martins}, {Kudryashova}, {Kull}, {Bachchan},
  {Lacoste-Seris}, {Lanza}, {Lavigne}, {Le Poncin-Lafitte}, {Lebreton},
  {Lebzelter}, {Leccia}, {Leclerc}, {Lecoeur-Taibi}, {Lemaitre}, {Lenhardt},
  {Leroux}, {Liao}, {Licata}, {Lindstr{\o}m}, {Lister}, {Livanou}, {Lobel},
  {L{\"o}ffler}, {L{\'o}pez}, {Lopez-Lozano}, {Lorenz}, {Loureiro},
  {MacDonald}, {Magalh{\~a}es Fernandes}, {Managau}, {Mann}, {Mantelet},
  {Marchal}, {Marchant}, {Marconi}, {Marie}, {Marinoni}, {Marrese},
  {Marschalk{\'o}}, {Marshall}, {Mart{\'\i}n-Fleitas}, {Martino}, {Mary},
  {Matijevi{\v{c}}}, {Mazeh}, {McMillan}, {Messina}, {Mestre}, {Michalik},
  {Millar}, {Miranda}, {Molina}, {Molinaro}, {Molinaro}, {Moln{\'a}r},
  {Moniez}, {Montegriffo}, {Monteiro}, {Mor}, {Mora}, {Morbidelli}, {Morel},
  {Morgenthaler}, {Morley}, {Morris}, {Mulone}, {Muraveva}, {Musella},
  {Narbonne}, {Nelemans}, {Nicastro}, {Noval}, {Ord{\'e}novic},
  {Ordieres-Mer{\'e}}, {Osborne}, {Pagani}, {Pagano}, {Pailler}, {Palacin},
  {Palaversa}, {Parsons}, {Paulsen}, {Pecoraro}, {Pedrosa}, {Pentik{\"a}inen},
  {Pereira}, {Pichon}, {Piersimoni}, {Pineau}, {Plachy}, {Plum}, {Poujoulet},
  {Pr{\v{s}}a}, {Pulone}, {Ragaini}, {Rago}, {Rambaux}, {Ramos-Lerate},
  {Ranalli}, {Rauw}, {Read}, {Regibo}, {Renk}, {Reyl{\'e}}, {Ribeiro},
  {Rimoldini}, {Ripepi}, {Riva}, {Rixon}, {Roelens}, {Romero-G{\'o}mez},
  {Rowell}, {Royer}, {Rudolph}, {Ruiz-Dern}, {Sadowski}, {Sagrist{\`a}
  Sell{\'e}s}, {Sahlmann}, {Salgado}, {Salguero}, {Sarasso}, {Savietto},
  {Schnorhk}, {Schultheis}, {Sciacca}, {Segol}, {Segovia}, {Segransan},
  {Serpell}, {Shih}, {Smareglia}, {Smart}, {Smith}, {Solano}, {Solitro},
  {Sordo}, {Soria Nieto}, {Souchay}, {Spagna}, {Spoto}, {Stampa}, {Steele},
  {Steidelm{\"u}ller}, {Stephenson}, {Stoev}, {Suess}, {S{\"u}veges}, {Surdej},
  {Szabados}, {Szegedi-Elek}, {Tapiador}, {Taris}, {Tauran}, {Taylor},
  {Teixeira}, {Terrett}, {Tingley}, {Trager}, {Turon}, {Ulla}, {Utrilla},
  {Valentini}, {van Elteren}, {Van Hemelryck}, {van Leeuwen}, {Varadi},
  {Vecchiato}, {Veljanoski}, {Via}, {Vicente}, {Vogt}, {Voss}, {Votruba},
  {Voutsinas}, {Walmsley}, {Weiler}, {Weingrill}, {Werner}, {Wevers},
  {Whitehead}, {Wyrzykowski}, {Yoldas}, {{\v{Z}}erjal}, {Zucker}, {Zurbach},
  {Zwitter}, {Alecu}, {Allen}, {Allende Prieto}, {Amorim},
  {Anglada-Escud{\'e}}, {Arsenijevic}, {Azaz}, {Balm}, {Beck}, {Bernstein},
  {Bigot}, {Bijaoui}, {Blasco}, {Bonfigli}, {Bono}, {Boudreault}, {Bressan},
  {Brown}, {Brunet}, {Bunclark}, {Buonanno}, {Butkevich}, {Carret}, {Carrion},
  {Chemin}, {Ch{\'e}reau}, {Corcione}, {Darmigny}, {de Boer}, {de Teodoro}, {de
  Zeeuw}, {Delle Luche}, {Domingues}, {Dubath}, {Fodor}, {Fr{\'e}zouls},
  {Fries}, {Fustes}, {Fyfe}, {Gallardo}, {Gallegos}, {Gardiol}, {Gebran},
  {Gomboc}, {G{\'o}mez}, {Grux}, {Gueguen}, {Heyrovsky}, {Hoar}, {Iannicola},
  {Isasi Parache}, {Janotto}, {Joliet}, {Jonckheere}, {Keil}, {Kim},
  {Klagyivik}, {Klar}, {Knude}, {Kochukhov}, {Kolka}, {Kos}, {Kutka}, {Lainey},
  {LeBouquin}, {Liu}, {Loreggia}, {Makarov}, {Marseille}, {Martayan},
  {Martinez-Rubi}, {Massart}, {Meynadier}, {Mignot}, {Munari}, {Nguyen},
  {Nordlander}, {Ocvirk}, {O'Flaherty}, {Olias Sanz}, {Ortiz}, {Osorio},
  {Oszkiewicz}, {Ouzounis}, {Palmer}, {Park}, {Pasquato}, {Peltzer}, {Peralta},
  {P{\'e}turaud}, {Pieniluoma}, {Pigozzi}, {Poels}, {Prat}, {Prod'homme},
  {Raison}, {Rebordao}, {Risquez}, {Rocca-Volmerange}, {Rosen}, {Ruiz-Fuertes},
  {Russo}, {Sembay}, {Serraller Vizcaino}, {Short}, {Siebert}, {Silva},
  {Sinachopoulos}, {Slezak}, {Soffel}, {Sosnowska}, {Strai{\v{z}}ys}, {ter
  Linden}, {Terrell}, {Theil}, {Tiede}, {Troisi}, {Tsalmantza}, {Tur},
  {Vaccari}, {Vachier}, {Valles}, {Van Hamme}, {Veltz}, {Virtanen}, {Wallut},
  {Wichmann}, {Wilkinson}, {Ziaeepour}, \& {Zschocke}}]{Gaia2016}
{Gaia Collaboration}, {Prusti}, T., {de Bruijne}, J.~H.~J., {et~al.} 2016,
  \aap, 595, A1, \dodoi{10.1051/0004-6361/201629272}

\bibitem[{{Gaia Collaboration} {et~al.}(2023){Gaia Collaboration}, {Vallenari},
  {Brown}, {Prusti}, {de Bruijne}, {Arenou}, {Babusiaux}, {Biermann},
  {Creevey}, {Ducourant}, {Evans}, {Eyer}, {Guerra}, {Hutton}, {Jordi},
  {Klioner}, {Lammers}, {Lindegren}, {Luri}, {Mignard}, {Panem}, {Pourbaix},
  {Randich}, {Sartoretti}, {Soubiran}, {Tanga}, {Walton}, {Bailer-Jones},
  {Bastian}, {Drimmel}, {Jansen}, {Katz}, {Lattanzi}, {van Leeuwen}, {Bakker},
  {Cacciari}, {Casta{\~n}eda}, {De Angeli}, {Fabricius}, {Fouesneau},
  {Fr{\'e}mat}, {Galluccio}, {Guerrier}, {Heiter}, {Masana}, {Messineo},
  {Mowlavi}, {Nicolas}, {Nienartowicz}, {Pailler}, {Panuzzo}, {Riclet}, {Roux},
  {Seabroke}, {Sordo}, {Th{\'e}venin}, {Gracia-Abril}, {Portell}, {Teyssier},
  {Altmann}, {Andrae}, {Audard}, {Bellas-Velidis}, {Benson}, {Berthier},
  {Blomme}, {Burgess}, {Busonero}, {Busso}, {C{\'a}novas}, {Carry}, {Cellino},
  {Cheek}, {Clementini}, {Damerdji}, {Davidson}, {de Teodoro}, {Nu{\~n}ez
  Campos}, {Delchambre}, {Dell'Oro}, {Esquej}, {Fern{\'a}ndez-Hern{\'a}ndez},
  {Fraile}, {Garabato}, {Garc{\'\i}a-Lario}, {Gosset}, {Haigron}, {Halbwachs},
  {Hambly}, {Harrison}, {Hern{\'a}ndez}, {Hestroffer}, {Hodgkin}, {Holl},
  {Jan{\ss}en}, {Jevardat de Fombelle}, {Jordan}, {Krone-Martins}, {Lanzafame},
  {L{\"o}ffler}, {Marchal}, {Marrese}, {Moitinho}, {Muinonen}, {Osborne},
  {Pancino}, {Pauwels}, {Recio-Blanco}, {Reyl{\'e}}, {Riello}, {Rimoldini},
  {Roegiers}, {Rybizki}, {Sarro}, {Siopis}, {Smith}, {Sozzetti}, {Utrilla},
  {van Leeuwen}, {Abbas}, {{\'A}brah{\'a}m}, {Abreu Aramburu}, {Aerts},
  {Aguado}, {Ajaj}, {Aldea-Montero}, {Altavilla}, {{\'A}lvarez}, {Alves},
  {Anders}, {Anderson}, {Anglada Varela}, {Antoja}, {Baines}, {Baker},
  {Balaguer-N{\'u}{\~n}ez}, {Balbinot}, {Balog}, {Barache}, {Barbato},
  {Barros}, {Barstow}, {Bartolom{\'e}}, {Bassilana}, {Bauchet}, {Becciani},
  {Bellazzini}, {Berihuete}, {Bernet}, {Bertone}, {Bianchi}, {Binnenfeld},
  {Blanco-Cuaresma}, {Blazere}, {Boch}, {Bombrun}, {Bossini}, {Bouquillon},
  {Bragaglia}, {Bramante}, {Breedt}, {Bressan}, {Brouillet}, {Brugaletta},
  {Bucciarelli}, {Burlacu}, {Butkevich}, {Buzzi}, {Caffau}, {Cancelliere},
  {Cantat-Gaudin}, {Carballo}, {Carlucci}, {Carnerero}, {Carrasco},
  {Casamiquela}, {Castellani}, {Castro-Ginard}, {Chaoul}, {Charlot}, {Chemin},
  {Chiaramida}, {Chiavassa}, {Chornay}, {Comoretto}, {Contursi}, {Cooper},
  {Cornez}, {Cowell}, {Crifo}, {Cropper}, {Crosta}, {Crowley}, {Dafonte},
  {Dapergolas}, {David}, {David}, {de Laverny}, {De Luise}, {De March}, {De
  Ridder}, {de Souza}, {de Torres}, {del Peloso}, {del Pozo}, {Delbo},
  {Delgado}, {Delisle}, {Demouchy}, {Dharmawardena}, {Di Matteo}, {Diakite},
  {Diener}, {Distefano}, {Dolding}, {Edvardsson}, {Enke}, {Fabre}, {Fabrizio},
  {Faigler}, {Fedorets}, {Fernique}, {Fienga}, {Figueras}, {Fournier},
  {Fouron}, {Fragkoudi}, {Gai}, {Garcia-Gutierrez}, {Garcia-Reinaldos},
  {Garc{\'\i}a-Torres}, {Garofalo}, {Gavel}, {Gavras}, {Gerlach}, {Geyer},
  {Giacobbe}, {Gilmore}, {Girona}, {Giuffrida}, {Gomel}, {Gomez},
  {Gonz{\'a}lez-N{\'u}{\~n}ez}, {Gonz{\'a}lez-Santamar{\'\i}a},
  {Gonz{\'a}lez-Vidal}, {Granvik}, {Guillout}, {Guiraud},
  {Guti{\'e}rrez-S{\'a}nchez}, {Guy}, {Hatzidimitriou}, {Hauser}, {Haywood},
  {Helmer}, {Helmi}, {Sarmiento}, {Hidalgo}, {Hilger}, {H{\l}adczuk}, {Hobbs},
  {Holland}, {Huckle}, {Jardine}, {Jasniewicz}, {Jean-Antoine Piccolo},
  {Jim{\'e}nez-Arranz}, {Jorissen}, {Juaristi Campillo}, {Julbe}, {Karbevska},
  {Kervella}, {Khanna}, {Kontizas}, {Kordopatis}, {Korn}, {K{\'o}sp{\'a}l},
  {Kostrzewa-Rutkowska}, {Kruszy{\'n}ska}, {Kun}, {Laizeau}, {Lambert},
  {Lanza}, {Lasne}, {Le Campion}, {Lebreton}, {Lebzelter}, {Leccia}, {Leclerc},
  {Lecoeur-Taibi}, {Liao}, {Licata}, {Lindstr{\o}m}, {Lister}, {Livanou},
  {Lobel}, {Lorca}, {Loup}, {Madrero Pardo}, {Magdaleno Romeo}, {Managau},
  {Mann}, {Manteiga}, {Marchant}, {Marconi}, {Marcos}, {Marcos Santos},
  {Mar{\'\i}n Pina}, {Marinoni}, {Marocco}, {Marshall}, {Martin Polo},
  {Mart{\'\i}n-Fleitas}, {Marton}, {Mary}, {Masip}, {Massari},
  {Mastrobuono-Battisti}, {Mazeh}, {McMillan}, {Messina}, {Michalik}, {Millar},
  {Mints}, {Molina}, {Molinaro}, {Moln{\'a}r}, {Monari}, {Mongui{\'o}},
  {Montegriffo}, {Montero}, {Mor}, {Mora}, {Morbidelli}, {Morel}, {Morris},
  {Muraveva}, {Murphy}, {Musella}, {Nagy}, {Noval}, {Oca{\~n}a}, {Ogden},
  {Ordenovic}, {Osinde}, {Pagani}, {Pagano}, {Palaversa}, {Palicio},
  {Pallas-Quintela}, {Panahi}, {Payne-Wardenaar}, {Pe{\~n}alosa Esteller},
  {Penttil{\"a}}, {Pichon}, {Piersimoni}, {Pineau}, {Plachy}, {Plum}, {Poggio},
  {Pr{\v{s}}a}, {Pulone}, {Racero}, {Ragaini}, {Rainer}, {Raiteri}, {Rambaux},
  {Ramos}, {Ramos-Lerate}, {Re Fiorentin}, {Regibo}, {Richards}, {Rios Diaz},
  {Ripepi}, {Riva}, {Rix}, {Rixon}, {Robichon}, {Robin}, {Robin}, {Roelens},
  {Rogues}, {Rohrbasser}, {Romero-G{\'o}mez}, {Rowell}, {Royer}, {Ruz Mieres},
  {Rybicki}, {Sadowski}, {S{\'a}ez N{\'u}{\~n}ez}, {Sagrist{\`a} Sell{\'e}s},
  {Sahlmann}, {Salguero}, {Samaras}, {Sanchez Gimenez}, {Sanna},
  {Santove{\~n}a}, {Sarasso}, {Schultheis}, {Sciacca}, {Segol}, {Segovia},
  {S{\'e}gransan}, {Semeux}, {Shahaf}, {Siddiqui}, {Siebert}, {Siltala},
  {Silvelo}, {Slezak}, {Slezak}, {Smart}, {Snaith}, {Solano}, {Solitro},
  {Souami}, {Souchay}, {Spagna}, {Spina}, {Spoto}, {Steele},
  {Steidelm{\"u}ller}, {Stephenson}, {S{\"u}veges}, {Surdej}, {Szabados},
  {Szegedi-Elek}, {Taris}, {Taylor}, {Teixeira}, {Tolomei}, {Tonello}, {Torra},
  {Torra}, {Torralba Elipe}, {Trabucchi}, {Tsounis}, {Turon}, {Ulla}, {Unger},
  {Vaillant}, {van Dillen}, {van Reeven}, {Vanel}, {Vecchiato}, {Viala},
  {Vicente}, {Voutsinas}, {Weiler}, {Wevers}, {Wyrzykowski}, {Yoldas}, {Yvard},
  {Zhao}, {Zorec}, {Zucker}, \& {Zwitter}}]{Gaia2023}
{Gaia Collaboration}, {Vallenari}, A., {Brown}, A.~G.~A., {et~al.} 2023, \aap,
  674, A1, \dodoi{10.1051/0004-6361/202243940}

\bibitem[{{Goodrich}(1989)}]{Goodrich1989}
{Goodrich}, R.~W. 1989, \apj, 342, 224, \dodoi{10.1086/167586}

\bibitem[{{Greene} \& {Ho}(2005)}]{Greene2005b}
{Greene}, J.~E., \& {Ho}, L.~C. 2005, \apj, 630, 122, \dodoi{10.1086/431897}

\bibitem[{{Greisen}(2003)}]{Greisen2003}
{Greisen}, E.~W. 2003, in Astrophysics and Space Science Library, Vol. 285,
  Information Handling in Astronomy - Historical Vistas, ed. A.~{Heck}, 109,
  \dodoi{10.1007/0-306-48080-8_7}

\bibitem[{{Gu} {et~al.}(2015){Gu}, {Chen}, {Komossa}, {Yuan}, {Shen}, {Wajima},
  {Zhou}, \& {Zensus}}]{Gu2015}
{Gu}, M., {Chen}, Y., {Komossa}, S., {et~al.} 2015, \apjs, 221, 3,
  \dodoi{10.1088/0067-0049/221/1/3}

\bibitem[{{Hardcastle} {et~al.}(1998){Hardcastle}, {Alexander}, {Pooley}, \&
  {Riley}}]{Hardcastle1998}
{Hardcastle}, M.~J., {Alexander}, P., {Pooley}, G.~G., \& {Riley}, J.~M. 1998,
  \mnras, 296, 445, \dodoi{10.1046/j.1365-8711.1998.01480.x}

\bibitem[{{Helfand} {et~al.}(2015){Helfand}, {White}, \&
  {Becker}}]{Helfand2015}
{Helfand}, D.~J., {White}, R.~L., \& {Becker}, R.~H. 2015, \apj, 801, 26,
  \dodoi{10.1088/0004-637X/801/1/26}

\bibitem[{{Hern{\'a}ndez-Garc{\'\i}a}
  {et~al.}(2017){Hern{\'a}ndez-Garc{\'\i}a}, {Panessa}, {Giroletti},
  {Ghisellini}, {Bassani}, {Masetti}, {Povi{\'c}}, {Bazzano}, {Ubertini},
  {Malizia}, \& {Chavushyan}}]{Hernandez-Garcia2017}
{Hern{\'a}ndez-Garc{\'\i}a}, L., {Panessa}, F., {Giroletti}, M., {et~al.} 2017,
  \aap, 603, A131, \dodoi{10.1051/0004-6361/201730530}

\bibitem[{{Hota} {et~al.}(2011){Hota}, {Sirothia}, {Ohyama}, {Konar}, {Kim},
  {Rey}, {Saikia}, {Croston}, \& {Matsushita}}]{Hota2011}
{Hota}, A., {Sirothia}, S.~K., {Ohyama}, Y., {et~al.} 2011, \mnras, 417, L36,
  \dodoi{10.1111/j.1745-3933.2011.01115.x}

\bibitem[{{Intema}(2014{\natexlab{a}})}]{Intema2014a}
{Intema}, H.~T. 2014{\natexlab{a}}, in Astronomical Society of India Conference
  Series, Vol.~13, Astronomical Society of India Conference Series, 469,
  \dodoi{10.48550/arXiv.1402.4889}

\bibitem[{{Intema}(2014{\natexlab{b}})}]{Intema2014b}
{Intema}, H.~T. 2014{\natexlab{b}}, {SPAM: Source Peeling and Atmospheric
  Modeling}, Astrophysics Source Code Library, record ascl:1408.006.
\newblock \doeprint{1408.006}

\bibitem[{{Intema} {et~al.}(2009){Intema}, {van der Tol}, {Cotton}, {Cohen},
  {van Bemmel}, \& {R{\"o}ttgering}}]{Intema2009}
{Intema}, H.~T., {van der Tol}, S., {Cotton}, W.~D., {et~al.} 2009, \aap, 501,
  1185, \dodoi{10.1051/0004-6361/200811094}

\bibitem[{{Ishwara-Chandra} {et~al.}(2020){Ishwara-Chandra}, {Taylor}, {Green},
  {Stil}, {Vaccari}, \& {Ocran}}]{Ishwara2020}
{Ishwara-Chandra}, C.~H., {Taylor}, A.~R., {Green}, D.~A., {et~al.} 2020,
  \mnras, 497, 5383, \dodoi{10.1093/mnras/staa2341}

\bibitem[{{J{\"a}rvel{\"a}} {et~al.}(2022){J{\"a}rvel{\"a}}, {Dahale},
  {Crepaldi}, {Berton}, {Congiu}, \& {Antonucci}}]{Jarvela2022}
{J{\"a}rvel{\"a}}, E., {Dahale}, R., {Crepaldi}, L., {et~al.} 2022, \aap, 658,
  A12, \dodoi{10.1051/0004-6361/202141698}

\bibitem[{{J{\"a}rvel{\"a}} {et~al.}(2018){J{\"a}rvel{\"a}},
  {L{\"a}hteenm{\"a}ki}, \& {Berton}}]{Jarvela2018}
{J{\"a}rvel{\"a}}, E., {L{\"a}hteenm{\"a}ki}, A., \& {Berton}, M. 2018, \aap,
  619, A69, \dodoi{10.1051/0004-6361/201832876}

\bibitem[{{J{\"a}rvel{\"a}} {et~al.}(2015){J{\"a}rvel{\"a}},
  {L{\"a}hteenm{\"a}ki}, \& {Le{\'o}n-Tavares}}]{Jarvela2015}
{J{\"a}rvel{\"a}}, E., {L{\"a}hteenm{\"a}ki}, A., \& {Le{\'o}n-Tavares}, J.
  2015, \aap, 573, A76, \dodoi{10.1051/0004-6361/201424694}

\bibitem[{{Jarvis} {et~al.}(2021){Jarvis}, {Harrison}, {Mainieri}, {Alexander},
  {Arrigoni Battaia}, {Calistro Rivera}, {Circosta}, {Costa}, {De Breuck},
  {Edge}, {Girdhar}, {Kakkad}, {Kharb}, {Lansbury}, {Molyneux}, {Mukherjee},
  {Mullaney}, {Farina}, {Silpa}, {Thomson}, \& {Ward}}]{Jarvis2021}
{Jarvis}, M.~E., {Harrison}, C.~M., {Mainieri}, V., {et~al.} 2021, \mnras, 503,
  1780, \dodoi{10.1093/mnras/stab549}

\bibitem[{{Jester} {et~al.}(2005){Jester}, {Schneider}, {Richards}, {Green},
  {Schmidt}, {Hall}, {Strauss}, {Vanden Berk}, {Stoughton}, {Gunn},
  {Brinkmann}, {Kent}, {Smith}, {Tucker}, \& {Yanny}}]{Jester2005}
{Jester}, S., {Schneider}, D.~P., {Richards}, G.~T., {et~al.} 2005, \aj, 130,
  873, \dodoi{10.1086/432466}

\bibitem[{{Jones} {et~al.}(2009){Jones}, {Read}, {Saunders}, {Colless},
  {Jarrett}, {Parker}, {Fairall}, {Mauch}, {Sadler}, {Watson}, {Burton},
  {Campbell}, {Cass}, {Croom}, {Dawe}, {Fiegert}, {Frankcombe}, {Hartley},
  {Huchra}, {James}, {Kirby}, {Lahav}, {Lucey}, {Mamon}, {Moore}, {Peterson},
  {Prior}, {Proust}, {Russell}, {Safouris}, {Wakamatsu}, {Westra}, \&
  {Williams}}]{Jones2009}
{Jones}, D.~H., {Read}, M.~A., {Saunders}, W., {et~al.} 2009, \mnras, 399, 683,
  \dodoi{10.1111/j.1365-2966.2009.15338.x}

\bibitem[{{Kaiser} {et~al.}(2000){Kaiser}, {Schoenmakers}, \&
  {R{\"o}ttgering}}]{Kaiser2000}
{Kaiser}, C.~R., {Schoenmakers}, A.~P., \& {R{\"o}ttgering}, H. J.~A. 2000,
  \mnras, 315, 381, \dodoi{10.1046/j.1365-8711.2000.03431.x}

\bibitem[{{Kellermann} {et~al.}(1989){Kellermann}, {Sramek}, {Schmidt},
  {Shaffer}, \& {Green}}]{Kellermann1989}
{Kellermann}, K.~I., {Sramek}, R., {Schmidt}, M., {Shaffer}, D.~B., \& {Green},
  R. 1989, \aj, 98, 1195, \dodoi{10.1086/115207}

\bibitem[{{Khamitov} {et~al.}(2022){Khamitov}, {Bikmaev}, {Gilfanov},
  {Sunyaev}, {Medvedev}, {Gorbachev}, \& {Irtuganov}}]{Khamitov2022}
{Khamitov}, I.~M., {Bikmaev}, I.~F., {Gilfanov}, M.~R., {et~al.} 2022,
  Astronomy Letters, 48, 724, \dodoi{10.1134/S1063773722110081}

\bibitem[{{Kharb} {et~al.}(2010{\natexlab{a}}){Kharb}, {Hota}, {Croston},
  {Hardcastle}, {O'Dea}, {Kraft}, {Axon}, \& {Robinson}}]{Kharb2010b}
{Kharb}, P., {Hota}, A., {Croston}, J.~H., {et~al.} 2010{\natexlab{a}}, \apj,
  723, 580, \dodoi{10.1088/0004-637X/723/1/580}

\bibitem[{{Kharb} {et~al.}(2010{\natexlab{b}}){Kharb}, {Lister}, \&
  {Cooper}}]{Kharb2010a}
{Kharb}, P., {Lister}, M.~L., \& {Cooper}, N.~J. 2010{\natexlab{b}}, \apj, 710,
  764, \dodoi{10.1088/0004-637X/710/1/764}

\bibitem[{{Kharb} {et~al.}(2006){Kharb}, {O'Dea}, {Baum}, {Colbert}, \&
  {Xu}}]{Kharb2006}
{Kharb}, P., {O'Dea}, C.~P., {Baum}, S.~A., {Colbert}, E.~J.~M., \& {Xu}, C.
  2006, \apj, 652, 177, \dodoi{10.1086/507945}

\bibitem[{{Kharb} {et~al.}(2008){Kharb}, {O'Dea}, {Baum}, {Daly}, {Mory},
  {Donahue}, \& {Guerra}}]{Kharb2008}
{Kharb}, P., {O'Dea}, C.~P., {Baum}, S.~A., {et~al.} 2008, \apjs, 174, 74,
  \dodoi{10.1086/520840}

\bibitem[{{King} {et~al.}(2017){King}, {Lohfink}, \& {Kara}}]{King2017}
{King}, A.~L., {Lohfink}, A., \& {Kara}, E. 2017, \apj, 835, 226,
  \dodoi{10.3847/1538-4357/835/2/226}

\bibitem[{{Kollatschny} \& {Zetzl}(2011)}]{Kollatschny2011}
{Kollatschny}, W., \& {Zetzl}, M. 2011, \nat, 470, 366,
  \dodoi{10.1038/nature09761}

\bibitem[{{Komatsu} {et~al.}(2011){Komatsu}, {Smith}, {Dunkley}, {Bennett},
  {Gold}, {Hinshaw}, {Jarosik}, {Larson}, {Nolta}, {Page}, {Spergel},
  {Halpern}, {Hill}, {Kogut}, {Limon}, {Meyer}, {Odegard}, {Tucker}, {Weiland},
  {Wollack}, \& {Wright}}]{Komatsu2011}
{Komatsu}, E., {Smith}, K.~M., {Dunkley}, J., {et~al.} 2011, \apjs, 192, 18,
  \dodoi{10.1088/0067-0049/192/2/18}

\bibitem[{{Komossa} {et~al.}(2006){Komossa}, {Voges}, {Xu}, {Mathur}, {Adorf},
  {Lemson}, {Duschl}, \& {Grupe}}]{Komossa2006}
{Komossa}, S., {Voges}, W., {Xu}, D., {et~al.} 2006, \aj, 132, 531,
  \dodoi{10.1086/505043}

\bibitem[{{Kukula} {et~al.}(1995){Kukula}, {Pedlar}, {Baum}, \&
  {O'Dea}}]{Kukula1995}
{Kukula}, M.~J., {Pedlar}, A., {Baum}, S.~A., \& {O'Dea}, C.~P. 1995, \mnras,
  276, 1262, \dodoi{10.1093/mnras/276.4.1262}

\bibitem[{{Lacy} {et~al.}(2020){Lacy}, {Baum}, {Chandler}, {Chatterjee},
  {Clarke}, {Deustua}, {English}, {Farnes}, {Gaensler}, {Gugliucci},
  {Hallinan}, {Kent}, {Kimball}, {Law}, {Lazio}, {Marvil}, {Mao}, {Medlin},
  {Mooley}, {Murphy}, {Myers}, {Osten}, {Richards}, {Rosolowsky}, {Rudnick},
  {Schinzel}, {Sivakoff}, {Sjouwerman}, {Taylor}, {White}, {Wrobel},
  {Andernach}, {Beasley}, {Berger}, {Bhatnager}, {Birkinshaw}, {Bower},
  {Brandt}, {Brown}, {Burke-Spolaor}, {Butler}, {Comerford}, {Demorest}, {Fu},
  {Giacintucci}, {Golap}, {G{\"u}th}, {Hales}, {Hiriart}, {Hodge}, {Horesh},
  {Ivezi{\'c}}, {Jarvis}, {Kamble}, {Kassim}, {Liu}, {Loinard}, {Lyons},
  {Masters}, {Mezcua}, {Moellenbrock}, {Mroczkowski}, {Nyland}, {O'Dea},
  {O'Sullivan}, {Peters}, {Radford}, {Rao}, {Robnett}, {Salcido}, {Shen},
  {Sobotka}, {Witz}, {Vaccari}, {van Weeren}, {Vargas}, {Williams}, \&
  {Yoon}}]{Lacy2020}
{Lacy}, M., {Baum}, S.~A., {Chandler}, C.~J., {et~al.} 2020, \pasp, 132,
  035001, \dodoi{10.1088/1538-3873/ab63eb}

\bibitem[{{Laor}(2000)}]{Laor2000}
{Laor}, A. 2000, \apjl, 543, L111, \dodoi{10.1086/317280}

\bibitem[{{Laor} \& {Behar}(2008)}]{Laor2008}
{Laor}, A., \& {Behar}, E. 2008, \mnras, 390, 847,
  \dodoi{10.1111/j.1365-2966.2008.13806.x}

\bibitem[{{Laor} {et~al.}(1994){Laor}, {Fiore}, {Elvis}, {Wilkes}, \&
  {McDowell}}]{Laor1994}
{Laor}, A., {Fiore}, F., {Elvis}, M., {Wilkes}, B.~J., \& {McDowell}, J.~C.
  1994, \apj, 435, 611, \dodoi{10.1086/174841}

\bibitem[{{Leighly}(1999{\natexlab{a}})}]{Leighly1999a}
{Leighly}, K.~M. 1999{\natexlab{a}}, \apjs, 125, 297, \dodoi{10.1086/313277}

\bibitem[{{Leighly}(1999{\natexlab{b}})}]{Leighly1999b}
---. 1999{\natexlab{b}}, \apjs, 125, 317, \dodoi{10.1086/313287}

\bibitem[{{Leighly} \& {Moore}(2004)}]{Leighly2004}
{Leighly}, K.~M., \& {Moore}, J.~R. 2004, \apj, 611, 107,
  \dodoi{10.1086/422088}

\bibitem[{{Lister} {et~al.}(2018){Lister}, {Aller}, {Aller}, {Hodge}, {Homan},
  {Kovalev}, {Pushkarev}, \& {Savolainen}}]{Lister2018}
{Lister}, M.~L., {Aller}, M.~F., {Aller}, H.~D., {et~al.} 2018, \apjs, 234, 12,
  \dodoi{10.3847/1538-4365/aa9c44}

\bibitem[{{Lister} {et~al.}(2013){Lister}, {Aller}, {Aller}, {Homan},
  {Kellermann}, {Kovalev}, {Pushkarev}, {Richards}, {Ros}, \&
  {Savolainen}}]{Lister2013}
---. 2013, \aj, 146, 120, \dodoi{10.1088/0004-6256/146/5/120}

\bibitem[{{Lister} {et~al.}(2016){Lister}, {Aller}, {Aller}, {Homan},
  {Kellermann}, {Kovalev}, {Pushkarev}, {Richards}, {Ros}, \&
  {Savolainen}}]{Lister2016}
---. 2016, \aj, 152, 12, \dodoi{10.3847/0004-6256/152/1/12}

\bibitem[{{Lister} {et~al.}(2019){Lister}, {Homan}, {Hovatta}, {Kellermann},
  {Kiehlmann}, {Kovalev}, {Max-Moerbeck}, {Pushkarev}, {Readhead}, {Ros}, \&
  {Savolainen}}]{Lister2019}
{Lister}, M.~L., {Homan}, D.~C., {Hovatta}, T., {et~al.} 2019, \apj, 874, 43,
  \dodoi{10.3847/1538-4357/ab08ee}

\bibitem[{{Marziani} {et~al.}(2018){Marziani}, {Dultzin}, {Sulentic}, {Del
  Olmo}, {Negrete}, {Mart{\'\i}nez-Aldama}, {D'Onofrio}, {Bon}, {Bon}, \&
  {Stirpe}}]{Marziani2018}
{Marziani}, P., {Dultzin}, D., {Sulentic}, J.~W., {et~al.} 2018, Frontiers in
  Astronomy and Space Sciences, 5, 6, \dodoi{10.3389/fspas.2018.00006}

\bibitem[{{Mathur} {et~al.}(2012){Mathur}, {Fields}, {Peterson}, \&
  {Grupe}}]{Mathur2012}
{Mathur}, S., {Fields}, D., {Peterson}, B.~M., \& {Grupe}, D. 2012, \apj, 754,
  146, \dodoi{10.1088/0004-637X/754/2/146}

\bibitem[{{Merloni} \& {Fabian}(2002)}]{Merloni2002}
{Merloni}, A., \& {Fabian}, A.~C. 2002, \mnras, 332, 165,
  \dodoi{10.1046/j.1365-8711.2002.05288.x}

\bibitem[{{Murgia} {et~al.}(2011){Murgia}, {Parma}, {Mack}, {de Ruiter},
  {Fanti}, {Govoni}, {Tarchi}, {Giacintucci}, \& {Markevitch}}]{Murgia2011}
{Murgia}, M., {Parma}, P., {Mack}, K.~H., {et~al.} 2011, \aap, 526, A148,
  \dodoi{10.1051/0004-6361/201015302}

\bibitem[{{Nagar} {et~al.}(1999){Nagar}, {Wilson}, {Mulchaey}, \&
  {Gallimore}}]{Nagar1999}
{Nagar}, N.~M., {Wilson}, A.~S., {Mulchaey}, J.~S., \& {Gallimore}, J.~F. 1999,
  \apjs, 120, 209, \dodoi{10.1086/313183}

\bibitem[{{Nandi} {et~al.}(2021){Nandi}, {Caproni}, {Kharb}, {Sebastian}, \&
  {Roy}}]{Nandi2021}
{Nandi}, S., {Caproni}, A., {Kharb}, P., {Sebastian}, B., \& {Roy}, R. 2021,
  \apj, 908, 178, \dodoi{10.3847/1538-4357/abd2ba}

\bibitem[{{Nandi} {et~al.}(2017){Nandi}, {Jamrozy}, {Roy}, {Larsson}, {Saikia},
  {Baes}, \& {Singh}}]{Nandi2017}
{Nandi}, S., {Jamrozy}, M., {Roy}, R., {et~al.} 2017, \mnras, 467, L56,
  \dodoi{10.1093/mnrasl/slw256}

\bibitem[{{Olgu{\'\i}n-Iglesias} {et~al.}(2020){Olgu{\'\i}n-Iglesias},
  {Kotilainen}, \& {Chavushyan}}]{Olguiniglesias2020}
{Olgu{\'\i}n-Iglesias}, A., {Kotilainen}, J., \& {Chavushyan}, V. 2020, \mnras,
  492, 1450, \dodoi{10.1093/mnras/stz3549}

\bibitem[{{Orban de Xivry} {et~al.}(2011){Orban de Xivry}, {Davies},
  {Schartmann}, {Komossa}, {Marconi}, {Hicks}, {Engel}, \&
  {Tacconi}}]{Orbandexivry2011}
{Orban de Xivry}, G., {Davies}, R., {Schartmann}, M., {et~al.} 2011, \mnras,
  417, 2721, \dodoi{10.1111/j.1365-2966.2011.19439.x}

\bibitem[{{Orr{\`u}} {et~al.}(2015){Orr{\`u}}, {van Velzen}, {Pizzo},
  {Yatawatta}, {Paladino}, {Iacobelli}, {Murgia}, {Falcke}, {Morganti}, {de
  Bruyn}, {Ferrari}, {Anderson}, {Bonafede}, {Mulcahy}, {Asgekar}, {Avruch},
  {Beck}, {Bell}, {van Bemmel}, {Bentum}, {Bernardi}, {Best}, {Breitling},
  {Broderick}, {Br{\"u}ggen}, {Butcher}, {Ciardi}, {Conway}, {Corstanje}, {de
  Geus}, {Deller}, {Duscha}, {Eisl{\"o}ffel}, {Engels}, {Frieswijk}, {Garrett},
  {Grie{\ss}meier}, {Gunst}, {Hamaker}, {Heald}, {Hoeft}, {van der Horst},
  {Intema}, {Juette}, {Kohler}, {Kondratiev}, {Kuniyoshi}, {Kuper}, {Loose},
  {Maat}, {Mann}, {Markoff}, {McFadden}, {McKay-Bukowski}, {Miley}, {Moldon},
  {Molenaar}, {Munk}, {Nelles}, {Paas}, {Pandey-Pommier}, {Pandey}, {Pietka},
  {Polatidis}, {Reich}, {R{\"o}ttgering}, {Rowlinson}, {Scaife},
  {Schoenmakers}, {Schwarz}, {Serylak}, {Shulevski}, {Smirnov}, {Steinmetz},
  {Stewart}, {Swinbank}, {Tagger}, {Tasse}, {Thoudam}, {Toribio}, {Vermeulen},
  {Vocks}, {van Weeren}, {Wijers}, {Wise}, \& {Wucknitz}}]{Orru2015}
{Orr{\`u}}, E., {van Velzen}, S., {Pizzo}, R.~F., {et~al.} 2015, \aap, 584,
  A112, \dodoi{10.1051/0004-6361/201526501}

\bibitem[{{Osterbrock} \& {Pogge}(1985)}]{Osterbrock1985}
{Osterbrock}, D.~E., \& {Pogge}, R.~W. 1985, \apj, 297, 166,
  \dodoi{10.1086/163513}

\bibitem[{{Panessa} {et~al.}(2019){Panessa}, {Baldi}, {Laor}, {Padovani},
  {Behar}, \& {McHardy}}]{Panessa2019}
{Panessa}, F., {Baldi}, R.~D., {Laor}, A., {et~al.} 2019, Nature Astronomy, 3,
  387, \dodoi{10.1038/s41550-019-0765-4}

\bibitem[{{Panessa} {et~al.}(2007){Panessa}, {Barcons}, {Bassani}, {Cappi},
  {Carrera}, {Ho}, \& {Pellegrini}}]{Panessa2007}
{Panessa}, F., {Barcons}, X., {Bassani}, L., {et~al.} 2007, \aap, 467, 519,
  \dodoi{10.1051/0004-6361:20066943}

\bibitem[{{Peterson}(2011)}]{Peterson2011}
{Peterson}, B.~M. 2011, in Narrow-Line Seyfert 1 Galaxies and their Place in
  the Universe, ed. L.~{Foschini}, M.~{Colpi}, L.~{Gallo}, D.~{Grupe},
  S.~{Komossa}, K.~{Leighly}, \& S.~{Mathur}, 32

\bibitem[{{Porcas}(2009)}]{Porcas2009}
{Porcas}, R.~W. 2009, \aap, 505, L1, \dodoi{10.1051/0004-6361/200912846}

\bibitem[{{Pounds} {et~al.}(1995){Pounds}, {Done}, \& {Osborne}}]{Pounds1995}
{Pounds}, K.~A., {Done}, C., \& {Osborne}, J.~P. 1995, \mnras, 277, L5,
  \dodoi{10.1093/mnras/277.1.L5}

\bibitem[{{Pringle}(1996)}]{Pringle1996}
{Pringle}, J.~E. 1996, \mnras, 281, 357, \dodoi{10.1093/mnras/281.1.357}

\bibitem[{{Pringle}(1997)}]{Pringle1997}
---. 1997, \mnras, 292, 136, \dodoi{10.1093/mnras/292.1.136}

\bibitem[{{Rakshit} {et~al.}(2018){Rakshit}, {Stalin}, {Hota}, \&
  {Konar}}]{Rakshit2018}
{Rakshit}, S., {Stalin}, C.~S., {Hota}, A., \& {Konar}, C. 2018, \apj, 869,
  173, \dodoi{10.3847/1538-4357/aaefe8}

\bibitem[{{Roos}(1988)}]{Roos1988}
{Roos}, N. 1988, \apj, 334, 95, \dodoi{10.1086/166820}

\bibitem[{{Saxton} {et~al.}(2008){Saxton}, {Read}, {Esquej}, {Freyberg},
  {Altieri}, \& {Bermejo}}]{Saxton2008}
{Saxton}, R.~D., {Read}, A.~M., {Esquej}, P., {et~al.} 2008, \aap, 480, 611,
  \dodoi{10.1051/0004-6361:20079193}

\bibitem[{{Schoenmakers} {et~al.}(2000){Schoenmakers}, {de Bruyn},
  {R{\"o}ttgering}, {van der Laan}, \& {Kaiser}}]{Schoenmakers2000}
{Schoenmakers}, A.~P., {de Bruyn}, A.~G., {R{\"o}ttgering}, H.~J.~A., {van der
  Laan}, H., \& {Kaiser}, C.~R. 2000, \mnras, 315, 371,
  \dodoi{10.1046/j.1365-8711.2000.03430.x}

\bibitem[{{Silpa} {et~al.}(2021){Silpa}, {Kharb}, {Harrison}, {Ho}, {Jarvis},
  {Ishwara-Chandra}, \& {Sebastian}}]{Silpa2021}
{Silpa}, S., {Kharb}, P., {Harrison}, C.~M., {et~al.} 2021, \mnras, 507, 991,
  \dodoi{10.1093/mnras/stab1870}

\bibitem[{{Singh} \& {Chand}(2018)}]{Singh2018}
{Singh}, V., \& {Chand}, H. 2018, \mnras, 480, 1796,
  \dodoi{10.1093/mnras/sty1818}

\bibitem[{{{\'S}niegowska} {et~al.}(2022){{\'S}niegowska}, {Panda}, {Czerny},
  {Savi{\'c}}, {Mart{\'\i}nez-Aldama}, {Marziani}, {Wang}, {Du}, {Popovi{\'c}},
  \& {Shekhar Saraf}}]{Sniegowska2022}
{{\'S}niegowska}, M., {Panda}, S., {Czerny}, B., {et~al.} 2022, arXiv e-prints,
  arXiv:2202.13839, \dodoi{10.48550/arXiv.2202.13839}

\bibitem[{{Sokolovsky} {et~al.}(2011){Sokolovsky}, {Kovalev}, {Pushkarev}, \&
  {Lobanov}}]{Sokolovsky2011}
{Sokolovsky}, K.~V., {Kovalev}, Y.~Y., {Pushkarev}, A.~B., \& {Lobanov}, A.~P.
  2011, \aap, 532, A38, \dodoi{10.1051/0004-6361/201016072}

\bibitem[{{Sridhar} {et~al.}(2020){Sridhar}, {Morganti}, {Nyland}, {Frank},
  {Harwood}, \& {Oosterloo}}]{Sridhar2020}
{Sridhar}, S.~S., {Morganti}, R., {Nyland}, K., {et~al.} 2020, \aap, 634, A108,
  \dodoi{10.1051/0004-6361/201936796}

\bibitem[{{Terashima} \& {Wilson}(2003)}]{Terashima2003}
{Terashima}, Y., \& {Wilson}, A.~S. 2003, \apj, 583, 145,
  \dodoi{10.1086/345339}

\bibitem[{{Thean} {et~al.}(2000){Thean}, {Pedlar}, {Kukula}, {Baum}, \&
  {O'Dea}}]{Thean2000}
{Thean}, A., {Pedlar}, A., {Kukula}, M.~J., {Baum}, S.~A., \& {O'Dea}, C.~P.
  2000, \mnras, 314, 573, \dodoi{10.1046/j.1365-8711.2000.03401.x}

\bibitem[{{Tingay}(1997)}]{Tingay1997}
{Tingay}, S.~J. 1997, \aap, 327, 550

\bibitem[{{Tingay} {et~al.}(1996){Tingay}, {Jauncey}, {Reynolds}, {Tzioumis},
  {Migenes}, {Gough}, {Lovell}, {McCulloch}, {Costa}, {Preston}, \&
  {Harbison}}]{Tingay1996}
{Tingay}, S.~J., {Jauncey}, D.~L., {Reynolds}, J.~E., {et~al.} 1996, \aj, 111,
  718, \dodoi{10.1086/117818}

\bibitem[{{Ulvestad} {et~al.}(2005){Ulvestad}, {Antonucci}, \&
  {Barvainis}}]{Ulvestad2005}
{Ulvestad}, J.~S., {Antonucci}, R.~R.~J., \& {Barvainis}, R. 2005, \apj, 621,
  123, \dodoi{10.1086/427426}

\bibitem[{{Urry} \& {Padovani}(1995)}]{Urry1995}
{Urry}, C.~M., \& {Padovani}, P. 1995, \pasp, 107, 803, \dodoi{10.1086/133630}

\bibitem[{{van Breugel} {et~al.}(1985){van Breugel}, {Miley}, {Heckman},
  {Butcher}, \& {Bridle}}]{VanBreugel1985}
{van Breugel}, W., {Miley}, G., {Heckman}, T., {Butcher}, H., \& {Bridle}, A.
  1985, \apj, 290, 496, \dodoi{10.1086/163007}

\bibitem[{{Vietri} {et~al.}(2022){Vietri}, {J{\"a}rvel{\"a}}, {Berton},
  {Ciroi}, {Congiu}, {Chen}, \& {Di Mille}}]{Vietri2022}
{Vietri}, A., {J{\"a}rvel{\"a}}, E., {Berton}, M., {et~al.} 2022, \aap, 662,
  A20, \dodoi{10.1051/0004-6361/202243523}

\bibitem[{{Wang} {et~al.}(2023){Wang}, {An}, {Guo}, {Ho}, {Baan}, {Braun},
  {Chen}, {Cheng}, {Hartley}, {Yang}, \& {Zhang}}]{Wang2023b}
{Wang}, A., {An}, T., {Guo}, S., {et~al.} 2023, \mnras, 523, L30,
  \dodoi{10.1093/mnrasl/slad051}

\bibitem[{{Wang} {et~al.}(2016){Wang}, {Du}, {Hu}, {Bai}, {Wang}, {Yi}, {Wang},
  {Zhang}, {Xin}, {Lun}, {Chang}, \& {Fan}}]{Wang2016}
{Wang}, F., {Du}, P., {Hu}, C., {et~al.} 2016, \apj, 824, 149,
  \dodoi{10.3847/0004-637X/824/2/149}

\bibitem[{{Wang} {et~al.}(1996){Wang}, {Brinkmann}, \& {Bergeron}}]{Wang1996}
{Wang}, T., {Brinkmann}, W., \& {Bergeron}, J. 1996, \aap, 309, 81

\bibitem[{{Webster} {et~al.}(2021{\natexlab{a}}){Webster}, {Croston},
  {Harwood}, {Baldi}, {Hardcastle}, {Mingo}, \&
  {R{\"o}ttgering}}]{Webster2021b}
{Webster}, B., {Croston}, J.~H., {Harwood}, J.~J., {et~al.} 2021{\natexlab{a}},
  \mnras, 508, 5972, \dodoi{10.1093/mnras/stab2939}

\bibitem[{{Webster} {et~al.}(2021{\natexlab{b}}){Webster}, {Croston}, {Mingo},
  {Baldi}, {Barkus}, {G{\"u}rkan}, {Hardcastle}, {Morganti}, {R{\"o}ttgering},
  {Sabater}, {Shimwell}, {Tasse}, \& {White}}]{Webster2021a}
{Webster}, B., {Croston}, J.~H., {Mingo}, B., {et~al.} 2021{\natexlab{b}},
  \mnras, 500, 4921, \dodoi{10.1093/mnras/staa3437}

\end{thebibliography}
\bibliographystyle{aasjournal}



\end{document}